\documentclass[12pt]{article}
\usepackage{amsmath,amssymb,amsbsy,amstext,amsthm,booktabs,simplewick,exscale,relsize,slashed,graphicx,amsfonts,upgreek}
\usepackage{multirow,dcolumn,bm,enumerate,url,hyperref,cite}

\newcommand{\gev}{{\rm GeV}}
\newcommand{\mev}{{\rm MeV}}

\title{\bf Supernovae Sparked By \\ \bf Dark Matter in White Dwarfs}

\author{Javier F.~Acevedo$^\eth$ and Joseph Bramante$^{\eth, \dagger}$\\
\small $^\eth$The Arthur B. McDonald Canadian Astroparticle Physics Research Institute, \\ \small Department of Physics, Engineering Physics, and Astronomy,\\ \small Queen's University, Kingston, Ontario, K7L 2S8, Canada\\
\small $^\dagger$Perimeter Institute for Theoretical Physics, Waterloo, Ontario, N2L 2Y5, Canada}

\begin{document}
\maketitle

\begin{abstract}
It was recently demonstrated that asymmetric dark matter can ignite supernovae by collecting and collapsing inside lone sub-Chandrasekhar mass white dwarfs, and that this may be the cause of Type Ia supernovae.  A ball of asymmetric dark matter accumulated inside a white dwarf and collapsing under its own weight, sheds enough gravitational potential energy through scattering with nuclei, to spark the fusion reactions that precede a Type Ia supernova explosion. In this article we elaborate on this mechanism and use it to place new bounds on interactions between nucleons and asymmetric dark matter for masses $m_{X} = 10^{6}-10^{16}$ GeV. Interestingly, we find that for dark matter more massive than $10^{11}$ GeV, Type Ia supernova ignition can proceed through the Hawking evaporation of a small black hole formed by the collapsed dark matter. We also identify how a cold white dwarf's Coulomb crystal structure substantially suppresses dark matter-nuclear scattering at low momentum transfers, which is crucial for calculating the time it takes dark matter to form a black hole. Higgs and vector portal dark matter models that ignite Type Ia supernovae are explored.
\end{abstract}

\tableofcontents

\section{Introduction}

While dark matter has been identified through its gravitational interactions in galaxies and the early universe, dark matter's mass, cosmological history, and non-gravitational interactions remain a mystery. Although the established evidence for dark matter is based on gravitational phenomena, it is reasonable to suppose that dark matter is coupled to known particles by more than just gravity. There is currently an extensive experimental effort to detect dark matter, including low-background underground experiments looking for dark matter bumping into Standard Model particles, space and ground-based searches for dark matter annihilating to Standard Model particles, and searches for dark matter in missing momentum events at particle colliders. 

Dark matter's non-gravitational interactions with ordinary matter may also be detected using astrophysical systems. The impact of dark matter interactions can be observed in solar fusion \cite{Press:1985ug,Gould:1987ju,Iocco:2012wk,Vincent:2014jia}, cooling gas cloud temperatures \cite{Chivukula:1989cc,Bhoonah:2018wmw,Bhoonah:2018gjb,Kovetz:2018zes}, stellar emission \cite{Raffelt:1996wa,An:2013yfc,Spolyar:2007qv}, white dwarfs \cite{Bramante:2015cua,Bertone:2007ae,McCullough:2010ai,Kouvaris:2010jy,Dreiner:2013tja,Leung:2013pra,Graham:2015apa,Graham:2018efk}, and neutron stars \cite{Goldman:1989nd,deLavallaz:2010wp,Kouvaris:2011fi,McDermott:2011jp,Bramante:2013hn,Bell:2013xk,Bramante:2014zca,Bramante:2015dfa,
Bramante:2016mzo,Bramante:2017ulk,Baryakhtar:2017dbj,Raj:2017wrv,Bell:2018pkk,Kouvaris:2018wnh,Ellis:2018bkr,Kouvaris:2015rea,Gresham:2018rqo}. 
In particular, due to the enormous density of white dwarfs and neutron stars, these objects serve as effective natural laboratories for testing dark matter models.

Dark matter may be the solution to some outstanding astrophysical puzzles. Contrary to prior expectations, recent evidence suggests that many Type Ia supernovae proceed from white dwarfs with masses below the Chandrasekhar mass threshold $M \lesssim 1.4M_{\odot}$ \cite{Scalzo:2014wxa,Scalzo:2018dai}. Reference \cite{Bramante:2015cua} proposed that asymmetric dark matter may be responsible for the ignition of Type Ia supernovae, including the ignition of sub-Chandrasekhar white dwarf progenitors (see \cite{Graham:2015apa,Graham:2018efk} for related analysis). Although other mechanisms have been proposed for the ignition of sub-Chandrasekhar white dwarfs, such as matter accretion from a neighbouring star \cite{1982ApJ...253..798N,Han:2003uj,Wang:2018pac}, binary mergers \cite{doi:10.1111/j.1365-2966.2011.19361.x,doi:10.1093/mnras/stw1575} or helium shell ignition \cite{2041-8205-770-1-L8}, these require binary companions, whereas some astronomical evidence indicates that Type Ia supernovae arise from lone white dwarfs \cite{2015Natur.521..332O,Maoz:2013hna}. 

The asymmetric dark matter ignition scenario proceeds as follows: first a large quantity of dark matter accumulates inside the white dwarf over time, by scattering against nuclei or electrons within it. As the dark matter settles inside the white dwarf after repeated scatters, it will thermalize at the white dwarf's internal temperature, leading to the formation of a thermalized dark matter sphere near the center of the white dwarf. This dark matter sphere grows in mass as more dark matter is accreted and thermalized; a matter sphere composed of asymmetric dark matter \cite{Petraki:2013wwa,Zurek:2013wia} will grow quickly, since asymmetric dark matter particles do not self-annihilate. Once such a dark matter sphere acquires a critical mass necessary for self-gravitation, its subsequent gravitational collapse releases enough energy through nuclear elastic scattering to dramatically heat the core of white dwarf stars. This heating prompts a thermonuclear runaway reaction that precedes a Type Ia supernova. The particular dynamics of dark matter accumulation and ignition have consequences for Type Ia supernova observations; intriguingly, \cite{Bramante:2015cua} showed that the observed inverse correlation \cite{Pan:2013cva,Scalzo:2018dai} between Type Ia host galaxy age and Type Ia light curve stretch (and by extension a correlation to white dwarf progenitor masses \cite{Scalzo:2018dai}), could be accounted for with asymmetric dark matter, which would tend to ignite less heavy white dwarfs at later times.

This document explores new aspects of the dark matter Type Ia supernova ignition mechanism. Using dark matter ignition of Type Ia supernovae, we set new bounds on the dark matter-nucleon cross section for heavy dark matter. These bounds depend on four relevant physical quantities: 1) the time it takes to accumulate a critical mass of dark matter in the center of the white dwarf, 2) the time it takes for the accreted dark matter to reach thermal equilibrium and settle to the center of the white dwarf, 3) the time it takes the dark matter sphere to collapse, 4) the energy transferred during dark matter collapse through scattering against stellar constituents. We set bounds on dark matter interactions by requiring that the first three processes take longer than the age of a known white dwarf, which has not exploded. To determine which white dwarf sets the most stringent bound, we have used the Montreal White Dwarf Database \cite{2017ASPC}, and found that white dwarf SDSS J160420.40+055542.3, a 3.3 Gyr old, $\sim 1.4 M_\odot$ mass white dwarf provides the strongest constraint; other white dwarf masses and ages are detailed, along with a proposal to use white dwarfs in old globular clusters to set more stringent bounds, in Appendix \ref{app:wds}.

Besides igniting Type Ia supernovae during collapse, we find that for a dark matter sphere which collapses without igniting the white dwarf, the black hole formed from the collapsed dark matter may itself ignite the white dwarf via Hawking radiation. The possibility that white dwarfs could be ignited via black hole evaporation was mentioned in Reference \cite{Graham:2018efk}, which studied the ignition of white dwarfs via Standard Model particle showers produced in the decay and annihilation of very massive dark matter particles in white dwarfs. Here we will find that in a similar fashion, small evaporating black holes formed from collapsed dark matter cores can ignite a white dwarf. This evaporative ignition will occur for dark matter heavier than $\sim 10^{11}$ GeV; such dark matter is heavy enough to collapse after a modest amount of dark matter has accumulated. This can result in the formation of a small black hole that evaporates within a few billion years. 

The evaporation of this black hole will lead to Type Ia supernova ignition via Hawking radiation, provided evaporation of the black hole occurs faster than accretion of stellar material and additional dark matter particles, because the black hole must become tiny enough for the Hawking radiation to reach temperatures suitable for ignition. For dark matter that ignites white dwarfs via black hole evaporation, the time that it takes for the dark matter sphere to collapse and form the black hole, is the limiting factor determining the constraint on dark matter interactions. We also find some region of the parameter space where the black hole formed grows, eventually leading to the destruction of the white dwarf, without producing a Type Ia supernova event -- such an eventuality for white dwarf stars was first considered in \cite{Kouvaris:2010jy}.

The structure of this paper is as follows: in Section \ref{sec:capture} we detail multiscatter capture of dark matter and discuss the formation of the dark matter sphere, its gravitational collapse, and the minimum cross section for this process to occur within the time required by observations. Next, we analyse the heating process as gravitational collapse proceeds, which yields new bounds on the dark matter-nucleon cross section. Our results can be compared to bounds from the Xenon-1T experiment \cite{Aprile:2017iyp} and neutron star implosions \cite{Bramante:2017ulk,Bramante:2015dfa}. Finally, we consider the black hole formation and ignition process via Hawking radiation, which yields new bounds for higher dark matter masses. As an example, we compare the bounds obtained to an explicit model for self-interacting dark matter and use them to set constraints on the model parameter space. In Section \ref{sec:conc}, we conclude. Throughout this paper, we work in natural units where $\hbar{=c=k_{b}=1}$ and $G=1/M_{pl}^{2}$ where $M_{pl} \approx 1.2 \times {10^{19} \ \gev}$ is the non-reduced Planck mass.

\section{Dark matter capture, thermalization and collapse in white dwarfs}
\label{sec:capture}

\subsection{Dark matter capture}
In order to have observable effects on a white dwarf star, a sizable amount of dark matter from the halo must become captured in its interior, by scattering and re-scattering with nuclei or electrons as the dark matter passes through the white dwarf. Here we will focus on dark matter capture via scattering with nuclei. The capture of dark matter in white dwarfs can be computed using methods outlined in \cite{Bramante:2017xlb}. In that work, a detailed analysis of the capture process is presented for the case that dark matter undergoes multiple scatters as it travels through the star. The multiscatter treatment is particularly relevant for dark matter masses in excess of $\sim 10^4$ GeV. Such a treatment is necessary for the present work, as we will consider heavy dark matter $m_{X}\geq{10^{6} \ \gev}$, which will require multiple scatters to slow down to sub-escape velocity and be captured. The total capture rate is 
\begin{equation}
 C_{X}=\sum_{N=1}^{\infty} C_{N}
 \label{cap1}
\end{equation}
Where each $C_{N}$ term in the sum corresponds to the capture rate for dark matter that underwent $N$ scatters before losing sufficient energy to become gravitationally bound to the white dwarf star. Note that, depending on the mass of the dark matter and its velocity in the halo, there will be a minimum number of scatters required in order to capture the dark matter. Each of these terms is given by
\begin{equation}
 C_{N}=\pi{R^{2}}p_{N}(\tau)\sqrt{\frac{6}{\pi}}\frac{n_{X}}{\bar{v}}\left[(2\bar{v}^{2}+3v_{esc}^{2})-(2\bar{v}^{2}+3v_{N}^{2})\exp{\left(-\frac{3(v_{N}^{2}-v_{esc}^{2})}{2\bar{v}^{2}}\right)}\right].
 \label{cap2}
\end{equation}
Here $R$ is the radius of the star, $n_{X}$ the dark matter halo number density, $v_{esc}$ is the escape velocity from the star surface ($v_{esc}\ll{1}$) and $\bar{v}$ is the average halo speed of dark matter. The quantity $v_{N}$ is defined as $v_{esc}(1-\beta_{+}/2)^{-N/2}$ with $\beta_{+}=4m_{X}m_{a}/(m_{X}+m_{a})^{2}$, and it accounts for the energy loss in successive momentum exchanges as a particle undergoes $N$ scatters. The function 
\begin{equation}
p_{N}(\tau) =  2 \int_0^1 dy~\frac{y e^{-y \tau} \left(y \tau \right)^N}{N!},
\end{equation} 
is a Poisson weighting that gives the probability of $N$ scatters in the star for stellar optical depth $\tau \equiv 3 \sigma_{aX} M / (2 \pi R^2 m_{a})$, where $y=\cos \theta$ defines the angle at which the dark matter entering the star and $M$ is the white dwarf mass, $\sigma_{aX}$ is the dark matter nuclear scattering cross-section, $R$ is the radius of the white dwarf, and $m_{a}$ is the mass of a nucleus in the white dwarf, see Ref.~\cite{Bramante:2017xlb}. The optical depth is proportional to the ratio between the dark matter-nuclei cross section and the saturation cross section for which a dark matter particle typically scatters about once as it traverses a travels through the star, $\sigma_{sat} \equiv \pi R^2 M/m_{a}$. We assume the fraction of energy lost in a collision follows a uniform distribution over $0<\Delta{E}/E_{0}<\beta_{+}$, where $E_{0}$ is the initial energy carried by the dark matter particle in the star's rest frame. Let us emphasize that, although the capture rate is proportional to the area projected by the star, it also depends on the optical depth which includes the information about interactions between dark matter and the stellar constituents and scales as $R^{-2}$. Therefore, multiscatter capture is more relevant for more compact stars.

Expression (\ref{cap2}) is valid for capture in white dwarfs; the exact capture rate depends on the white dwarf's composition. Here we conservatively assume the inner composition of white dwarfs is predominantly $^{12}$C nuclei. Dark matter possessing spin-independent interactions ($e.g.$ a vector or scalar current) will coherently scatter off the nucleons within these nuclei for low energy exchanges. In contrast, at higher energies and momentum exchange, the scattering cross-section becomes suppressed as the dark matter begins to scatter with individual nucleons. To account for this, we make the following substitution for the dark matter-nuclear cross section, which determines the optical depth $\tau$,
\begin{equation}
\sigma_{aX} \  \rightarrow \ \sigma'_{aX}=A^2 \left(\frac{\mu_{aX}}{\mu_{nX}}\right)^{2}S(q)F^{2}(\left\langle{E_{R}}\right\rangle) \ \sigma_{nX},
\label{eq:sax}
\end{equation}
where $\sigma_{aX}$ is the dark matter-nuclear cross section for a nucleus with $A$ nucleons, written in terms of the dark matter-nucleon cross section $\sigma_{nX}$ (here assumed to be velocity-independent), $\mu_{aX}$ is the dark matter-nuclear reduced mass, and $\mu_{nX}$ is the dark matter-nucleon reduced mass. The function $S(q)$ is the static structure factor that accounts for correlation effects in an ion lattice, and depends on the momentum transfer $q \approx m_{a} \sqrt{2E/m_{X}}$, in the limit $m_{a} \ll m_{X}$. For the recoil energies involved during capture, however, it evaluates to unity. This structure factor becomes relevant for both thermalization and gravitational collapse of the captured dark matter, which we discuss below. The second factor $F^{2}(\left\langle{E_{R}}\right\rangle)$ is the Helm form factor \cite{Primack:1988zm,PhysRev.104.1466,LEWIN199687} evaluated at the average recoil energy 
\begin{align}
\left\langle{E_{R}}\right\rangle \equiv \frac{\int_{0}^{E_{R}^{max}}E \ F^{2}(E) \ dE}{\int_{0}^{E_{R}^{max}}F^{2}(E) \ dE} \approx{1 \ \mev}
\end{align}
for heavy dark matter scattering off $^{12}$C at velocity $v_{esc} \approx 10^{-2}$, making the average Helm form factor correction of order $\sim 0.5$.  Momentum exchanges with recoil energies much in excess of $\sim$ MeV are suppressed by this form factor. To account for the Helm form factor in the energy loss expression, we replace the factor of 1/2 that multiplies $\beta_{+}$ in $v_{N}$ by $\left\langle{E_{R}}\right\rangle/E_{R}^{Max}$, where $E_{R}^{Max}=4m_{a}v_{esc}^{2}$ is the maximum recoil energy. Applying these corrections to (\ref{cap2}), and truncating (\ref{cap1}) at a sufficiently large value of $N$ such that the higher order terms removed are negligible, we arrive at the capture rate for heavy dark matter passing through white dwarfs. While we use the full capture rate \eqref{cap2} in calculations throughout this document, for ease of reference, the approximate dark matter mass capture rate on a white dwarf with mass $M=1.4~{M_{\odot}}$ and radius $R=2500 ~{\rm km}$, composed entirely of $^{12}$C, for dark matter masses well in excess of $m_X =10^4$ GeV (see \cite{Bramante:2017xlb}) is given by
\begin{align}
\dot{m}_X \approx {\rm Min } &\left[ 3 \times 10^{27}~{\frac{\rm GeV}{\rm s}}, 6 \times 10^{24}~{\frac{\rm GeV}{\rm s}}~\left( \frac{10^8~{\rm GeV}}{m_X} \right) \left( \frac{\sigma_{nX}}{10^{-42} ~{\rm cm^2}} \right) \right] \nonumber \\ &\times \left( \frac{\rho_{X}}{0.3~{\rm GeV/cm^3}} \right) \left( \frac{10^{-3}}{\bar{v}} \right) ,
\label{eq:capturedm-approx}
\end{align}
where this expression assumes a conservatively low, $2 \sigma$ allowed dark matter halo density of $\rho_X \approx 0.3~{\rm GeV/cm^3}$ \cite{Benito:2019ngh}, and a characteristic dark matter velocity of $\bar{v} \approx 10^{-3}$.

\subsection{Dark Matter Thermalization}
After it becomes captured and gravitationally bound, a dark matter particle will orbit through the white dwarf star, losing more kinetic energy in successive scatters. Thermalization of the captured dark matter occurs in two stages -- a stage where the dark matter orbits outside the star and a stage where it orbits entirely inside the star. Some aspects of dark matter thermalization inside a white dwarf will differ from dark matter thermalization in a main sequence star, studied in \cite{Kouvaris:2010jy}.

\subsection{First Thermalization}
In the initial thermalization stage, the dark matter particles follow a closed orbit that intersects the white dwarf, losing kinetic energy through scattering while crossing through the white dwarf. Upon capture, assuming a radial path through the star, a dark matter particle will have a turning point $r$ as a function of energy given by
\begin{equation}
 r=-\frac{GMm_{X}}{E}
 \label{eq:deltae1}
\end{equation}
Where $E<0$ is the particle's initial energy after capture. The average energy loss when a dark matter particle scatters against a nucleus at a distance $\tilde{r}$ from the star center, in the kinematic limit $m_{a}\ll{m_{X}}$, will be 
\begin{equation}
 \Delta{E}=2\frac{m_{a}}{m_{X}}\left(E+GMm_{X}\left(\frac{3R^{2}-\tilde{r}^{2}}{2R^{3}}\right)\right)
\end{equation}
Where the whole factor in parenthesis is the dark matter kinetic energy, $i.e.$ it is the sum of the particle's initial energy $E$ and the potential energy inside the star. To obtain the average energy loss for single scatter, we express $E$ in terms of the turning point $r$ using \eqref{eq:deltae1} and average over the star size
\begin{equation}
\Delta{E_{tr}}=2GMm_{a}\left(-\frac{1}{r}-\frac{1}{R}\int_{0}^{R}\frac{3R^{2}-\tilde{r}^{2}}{2R^{3}} \ d\tilde{r}\right)
\end{equation}
\begin{align}
\Delta E_{tr} =2GMm_{a} \left(\frac{4}{3R} -\frac{1}{r} \right).
\label{eq:deltae2}
\end{align}
The time it takes the dark matter to complete one orbit through the white dwarf is approximately the orbital period for a dark matter-white dwarf orbital radius (semi-major axis) $r$, 
\begin{equation}
\Delta t_{tr} = 2 \pi r^{3/2}/\sqrt{M},
\label{eq:deltat}
\end{equation}
where the initial size of dark matter's orbit around the white dwarf, $r_0$ can be determined from the captured dark matter's initial kinetic energy, $r_0 \approx \frac{\bar{v}^2}{2 GM}$. Finally, the average energy loss during the first thermalization phase is obtained by dividing Eq.~\eqref{eq:deltae2} by Eq.~\eqref{eq:deltat} and multiplying by the white dwarf optical depth $\tau$ to estimate the average number of times that the dark matter scatters as it completes a single transit through the white dwarf. 
\begin{align}
\frac{dE_{tr}}{dt_{tr}} = \frac{(GM)^{3/2} \tau m_{a}}{\pi r^{3/2}} \left( \frac{4}{3R} -\frac{1}{r}\right).
\end{align}
It is useful to make the variable substitution $\epsilon = R/r$, where this is approximately the ratio of the dark matter's initial kinetic energy $GMm_{X}/r$ to its binding energy at the surface of the star $GMm_X/R$. The time for the dark matter to settle inside the star after repeated scatters is
\begin{align}
t_1^{th} = \frac{\pi m_X R^{3/2}}{\sqrt{GM}  m_{a}} \int_{\epsilon_0}^{\epsilon_f} \frac{d \epsilon}{\tau \epsilon^{3/2}\left(\frac{4}{3}-\epsilon \right)}, 
\end{align}
where the initial ratio is $\epsilon_0 \approx R/r_0$ and the final ratio is $\epsilon_f \approx 1$ when the dark matter lies inside the star. Note that in the above expression the optical depth $\tau$, and by extension the dark matter-nuclear cross section $\sigma_{aX}$ depend on energy, because of the Helm form factor shown in Eq.~\eqref{eq:sax}. Therefore, we have conservatively taken the Helm form factor in $\tau(\sigma_{aX})$ to be $F^{2}(\left\langle{E_{R}}\right\rangle) \sim 0.5$, which is its minimum value for dark matter-nuclear recoil energies during the first stage of thermalization. The recoil energies involved are still too high in this early thermalization stage for the structure factor $S(q)$ to become relevant, we leave its discussion for the following section. Inserting typical white dwarf parameters, the time to complete the first stage of thermalization, after which the dark matter particle will orbit entirely inside the star, is given by
\begin{equation}
 t_{1}^{th} \lesssim {10^{-4} \ {\rm yrs} \left(\frac{10^{-40} \ {\rm cm^{2}}}{\sigma_{nX}}\right) \left(\frac{m_{X}}{10^{6} \ \gev}\right) \left(\frac{1.4{\rm M_{\odot}}}{M}\right)^{\frac{3}{2}}\left(\frac{R}{2500 \rm \ {\rm km}}\right)^{\frac{7}{2}}}.
 \label{ttherm1}
\end{equation}

\subsection{Second Thermalization}
During a second thermalization stage, the dark matter particles orbit entirely inside the white dwarf. As the dark matter particles move through the star interior, repeated scattering shrinks their orbits to a ``thermal radius" $r_{th}$ -- at this radius they have a thermal velocity distribution determined by the white dwarf core temperature $T$. We will now see that the structure of the star plays a critical role in this process.

At standard densities and temperatures for old massive ($\gtrsim M_\odot$) white dwarfs, atoms are completely ionized and electrons can be treated as a uniform free gas providing a neutralizing background for nuclei. Such a plasma, under high pressure and low temperature, solidifies to form a Coulomb crystal. We are considering old white dwarfs that have not exploded, with core temperatures $T \sim {10^{7}} \ \rm{K}$ that lie below the melting temperature of the plasma $T_{m} \sim 5 \times 10^{8} \ \rm{K}$ \cite{1993ApJ...414..695C,PhysRevLett.76.4572}. Therefore, we must consider a structure factor that accounts for dark matter scattering in a dense crystalline medium \cite{kittel1987quantum}. 

The structure factor of Coulomb crystals has been calculated in \cite{Baiko:1998xk,PhysRevE.61.1912}. It is normally split into two contributions corresponding to elastic Bragg scattering, and inelastic absorption or emission of phonons,
\begin{equation}
 S(q)=S'(q)+S''(q)
 \label{eq:struct}
\end{equation}
For completeness, we first present the elastic part $S'(q)$, which will not be essential for our analysis. The elastic part is written as \cite{kittel1987quantum},
\begin{equation}
 S'(q)= e^{-W(q)} (2\pi)^{3} n_{a} \sum_{\mathbf{G\neq 0}}\delta (\mathbf{q-G}),
\end{equation}
where $\mathbf{G}$ represents the reciprocal lattice vectors, $i.e.$ all nonzero vectors pointing between nuclei, $n_{a}=\rho_{wd}/m_{a}$ is the nucleus number density and $W(q)$ is the Debye-Waller factor. The Debye-Waller factor, which as we will see also appears in the inelastic structure factor, can be approximated by $W(q) \approx\langle{r_{T}^{2}}\rangle q^{2}/3$, where $\langle{r_{T}^{2}}\rangle$ is the mean-squared nucleus displacement at a given temperature $T$ \cite{kittel1987quantum}. We can estimate the mean-squared displacement by assuming each nucleus is bound by a harmonic oscillator potential, with its natural frequency given by the plasma frequency $\Omega_{p}^{2}=4 \pi \rho_{wd} Z^{2} e^{2}/m_{a}^{2}$, and summing over all normal modes per ion \cite{Shapiro:1983du},
\begin{equation}
 \langle{r_{T}^{2}}\rangle \approx \frac{14T}{m_{a}\Omega_{p}^{2}}
\end{equation}
Using $q^{2}=2m_{a}^{2}E/m_{X}$, we conveniently rewrite $W(q) = E/E_{sup}$, where
\begin{align}
E_{sup}\approx \frac{Z^{2}e^{2}m_{X}\rho_{wd}}{Tm_{a}^{3}}
\end{align} is the approximate energy scale at which dark matter nuclear scattering becomes suppressed. The Bragg scattering contribution $S'(q)$ evaluates exactly to zero when the momentum transfer roughly falls below the inverse nuclear spacing $a^{-1}=(4\pi\rho_{wd}/3m_{a})^{1/3}$. The sharp peaks of $S'(q)$ have a height $\lesssim 1$ close to this threshold value and quickly decay for higher momentum transfers due to the exponential suppression $e^{-W(q)}$. Because the second thermalization time will be predominantly determined by inelastic phonon scattering, the precise form of the elastic Bragg component, and its peaks around $q \sim |\mathbf{G}|$, will have a negligible contribution to the computations that follow.

The inelastic, phonon emission and absorption portion of the structure factor can be approximated by 
\begin{equation}
 S''(q) \approx 1-e^{-W(q)}=1-e^{-E/E_{sup}}.
 \label{eq:S''}
\end{equation}
This approximation is accurate when $q\gtrsim{3/a}$ \cite{PhysRevE.61.1912}. It also reproduces correct limits such that $S(q)\rightarrow 0$ and $S(q)\rightarrow 1$ when $q\rightarrow 0$ and $q\rightarrow \infty$ respectively. To our knowledge, this structure factor has not been accounted for in any previous work considering dark matter scattering with nuclei in white dwarfs. Here, to get a first estimate, we will exclusively apply Eq.~\eqref{eq:S''} to account for low momentum suppression of scattering, since in the high momentum exchange limit $S(q) \approx 1$, and can be safely neglected. On the other hand, for low momentum exchange we will find that the inclusion of the $S''(q)$ phonon structure factor results in much longer thermalization times, since momentum transfer to nuclei decreases and becomes comparable to the nuclear spacing as dark matter thermalizes. 

To compute the energy loss rate, first we consider the dark matter-nucleus interaction time $t_{aX}$,
\begin{equation}
t_{aX}=\frac{1}{n_{a}\sigma_{aX} v_{rel}},
\end{equation}
and the average energy loss per scatter, in the limit that $m_{a}\ll{m_{X}}$,
\begin{equation}
 \Delta{E}=\frac{2m_{a}}{m_{X}}E
\end{equation}
where $E$ is the energy of the dark matter particle and $v_{rel}$ is the relative velocity of dark matter-nucleus system. Thus, the energy loss rate is expressed as
\begin{equation}
 \frac{dE}{dt}=\frac{2\rho_{wd}\sigma_{aX}v_{rel}E}{m_{X}}.
\end{equation}

After completing the first thermalization phase, dark matter particles will have an initial energy given by the binding energy at the surface of the star $\sim GMm_{X}/R$. For energies above $E_{sup}$, they are in an inertial regime where $v_{rel} \sim \sqrt{2E/m_{X}}$, $i.e.$ their velocity dominates the scattering time $t_{aX}$ \cite{Janish:2019nkk}. In such inertial regime, the energy loss rate is
\begin{equation}
 \left(\frac{dE}{dt}\right)_{in}=\frac{\rho_{wd}A^{4}(1-e^{-E/E_{sup}})\sigma_{nX}}{m_{X}}\sqrt{\frac{2E}{m_{X}}}E \approx\frac{\sqrt{2}\rho_{wd}A^{4}\sigma_{nX}E^{3/2}}{m_{X}^{3/2}}
 \label{eq:th-inertial}
\end{equation}
In the above expression, we have conservatively set $F^{2}(\left\langle{E_{R}}\right\rangle) \sim 0.5$ for the recoil energies involved. When the energy drops below $E_{sup}$, dark matter particles enter the viscous regime where scattering rate is now dominated by the thermal speed of nuclei, which we estimate as $v_{rel} \sim v_{nuc}\sim \sqrt{T/m_{a}}$. In this new viscous regime, the energy loss rate is
\begin{equation}
 \left(\frac{dE}{dt}\right)_{vis}=\frac{2\rho_{wd}A^{4}(1-e^{-E/E_{sup}})\sigma_{nX}}{m_{X}}\sqrt{\frac{T}{m_{a}}}E\approx\frac{2\rho_{wd}A^{4}\sigma_{nX}\sqrt{T}E^{2}}{m_{X}\sqrt{m_{a}}E_{sup}},
 \label{eq:th-vis}
\end{equation}
where now the Helm form factor can be set to unity for the low recoil momenta. Throughout the rest of this paper, we shall assume $m_{a}\approx{12 \ \gev}$, $A=12$ and $Z=6$ for $^{12}$C nuclei. After entering the viscous regime with an approximate energy $E_{sup}$, the dark matter energy is reduced over time to the thermal value $E\approx{3T/2}$. We find that the time required for dark matter to reach this thermal energy in the viscous regime to be several orders of magnitude larger than the time spent in the inertial regime. Such time is estimated by integrating \eqref{eq:th-vis},
\begin{align}
 t_{2}^{th}&=\frac{m_{X}\sqrt{m_{a}}E_{sup}}{2\rho_{wd}A^{4}\sigma_{nX}\sqrt{T}}\int_{E_{sup}}^{3T/2}{\frac{dE}{E^{2}}}\approx{\frac{m_{X}E_{sup}\sqrt{m_{a}}}{3T^{3/2}\rho_{wd}A^{4}\sigma_{nX}}} \nonumber \\ & \approx{20 \ {\rm yrs} \left(\frac{10^{-40} \ {\rm cm^{2}}}{\sigma_{nX}}\right)\left(\frac{m_{X}}{10^{6} \ \gev}\right)^{2}\left(\frac{10^{7} \ \rm K}{T}\right)^{\frac{5}{2}}}
\end{align}
We remark that for typical white dwarf parameters and dark matter masses considered in this work, the time to complete the first phase of thermalization is negligible compared to the second phase. The minimum cross section for completing the first and second thermalization stages within 3 Gyrs are plotted in Figure \ref{fig:therm}. 

\subsection{Dark Matter Collapse}
After being captured and thermalized through repeated scatterings with white dwarf particles, the dark matter particles settle into a gravitationally bound configuration. The radius of this thermalized sphere of dark matter in the center of the white dwarf, $r_{th}$, can be computed using the virial theorem, equating the dark matter thermal energy at temperature $T$ with the dark matter's gravitational potential energy inside a white dwarf with density $\rho_{wd}$
\begin{equation}
r_{th}=\sqrt{\frac{9T}{4\pi G \rho_{wd} m_{X}}}\approx{30  {\rm~ m} \left(\frac{10^{6} \ \gev}{m_{X}}\right)^{\frac{1}{2}}  \left(\frac{10^{9} {\rm ~ \frac{g}{cm^3}}}{\rho_{wd}}\right)^{\frac{1}{2}}\left(\frac{T}{10^7 \ \rm K}\right)^{\frac{1}{2}}}
\end{equation}
Unless the dark matter is asymmetric \cite{Petraki:2013wwa,Zurek:2013wia} or equivalently has very small self-annihilation interactions \cite{Bramante:2013hn} the dark matter sphere will be prohibited from substantially growing by annihilation. Therefore, throughout this work we will consider asymmetric dark matter that does not have self-annihilation interactions.

Dark matter collected into a sphere of radius $r_{th}$ at the center of the white dwarf will collapse under its own weight, so long as two criteria are met: 
\begin{enumerate}
\item Dark matter must be self-gravitating, so that its gravitational binding energy decreases during collapse. If the gravitational potential of the white dwarf is much larger than the dark matter's self-gravity, then the binding energy of dark matter with mass $M_{dm}$ has the form of a harmonic oscillator potential, $E \sim - G M_{dm} \rho_{wd} r_{th}^2$. Under the influence of the white dwarf's gravitational potential alone, the dark matter sphere's potential energy would increase as the dark matter sphere shrank in size. As a consequence, without self-gravitation, the minimum energy configuration of the dark matter sphere will be at radius $r_{th}$, and collapse will not occur. Therefore, we require the density of the dark matter sphere to roughly exceed the white dwarf density, in order for collapse to occur.
\item The dark matter sphere must satisfy a Jeans instability condition.
\end{enumerate}

\begin{figure}[t] 
\includegraphics[scale=0.7]{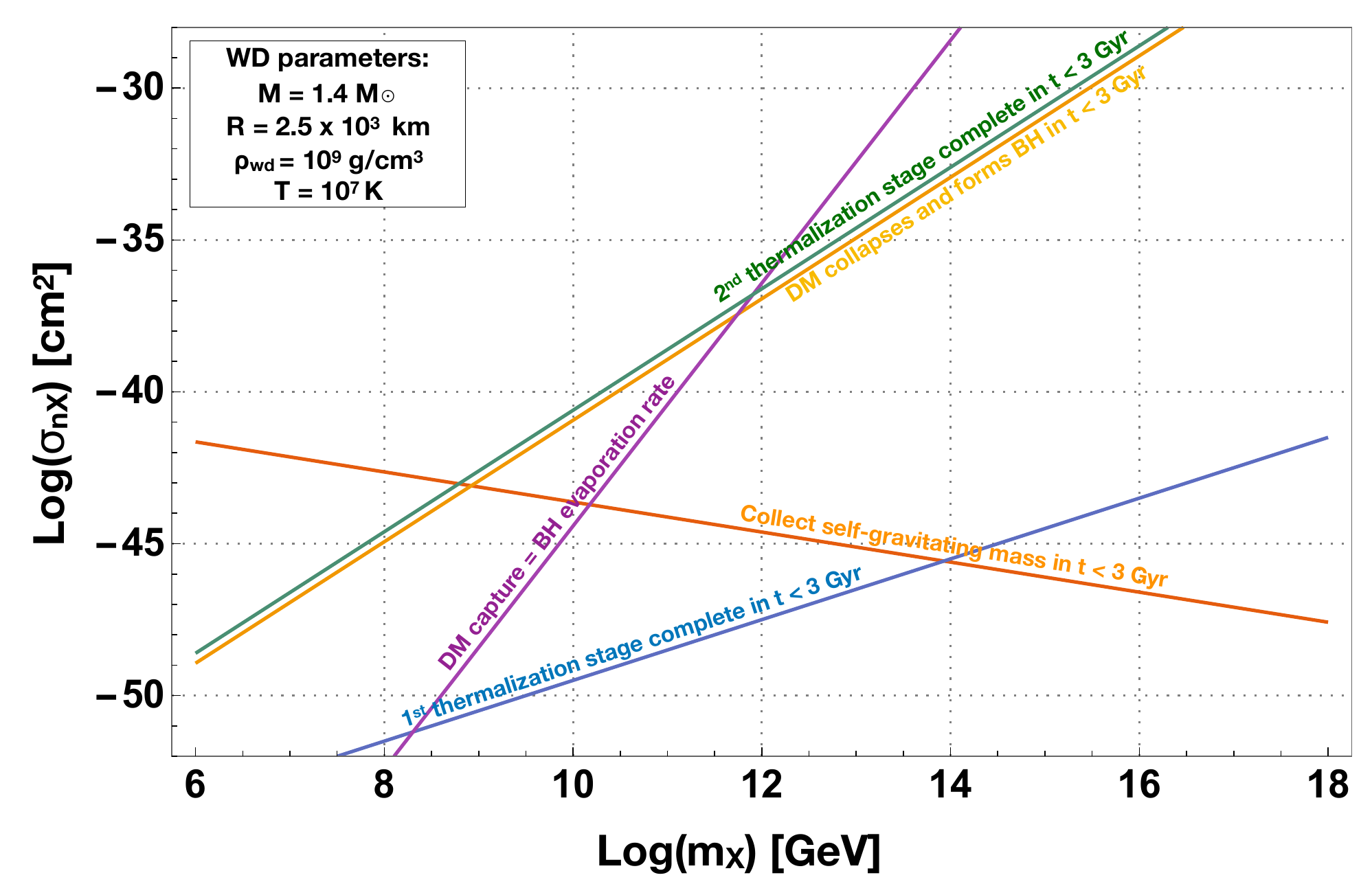} 
\caption{Minimum dark matter-nucleon cross section $\sigma_{nX}$ as a function of $m_{X}$, in a log-log plot, for dark matter to thermalize with the white dwarf (green and blue) and accumulate critical mass for self-gravitation (orange) within 3 Gyrs. The minimum cross section for the thermalized dark matter sphere to collapse to a black hole (yellow) within 3 Gyrs, and for dark matter accumulation to preclude complete black hole evaporation (purple) are also shown. A dark matter halo density $\rho_{X}\approx 0.3 \ \rm{GeV/cm^{3}}$ at the location of the white dwarf is assumed. The upper-left box shows the white dwarf parameters used, where the mass and age of the white dwarf were determined by the oldest massive white dwarf in the Montreal White Dwarf Database \cite{2017ASPC}, SDSS J160420.40+055542.3. }
\label{fig:therm}
\end{figure}

The first self-gravitation condition is met so long as the dark matter density equals or exceeds the white dwarf density within the thermalization radius, $r_{th}$. This defines a critical self-gravitating mass $M_{crit} = 4 \pi \rho_{wd} r_{th}^3 /3$. The number of dark matter particles required is
\begin{equation}
 N_{sg}=\frac{M_{crit}}{m_{X}}\approx 6\times{10^{37}} \left(\frac{10^{6} \ \gev}{m_{X}}\right)^{\frac{5}{2}} \left(\frac{10^{9} {\rm \frac{g}{cm^3}}}{\rho_{wd}}\right)^{\frac{1}{2}} \left(\frac{T}{10^7 \ {\rm K}}\right)^{\frac{3}{2}}.
 \label{eq:nsg}
\end{equation}
In Figure \ref{fig:therm}, the minimum dark matter-nucleon cross section required to collect a self-gravitating clump of dark matter in a white dwarf in 3 Gyrs is shown.

To determine whether the dark matter sphere is unstable, we perform a simple Jeans mass analysis, requiring that the sound-crossing time exceed the free-fall time of the dark matter sphere. The free-fall time is given by
\begin{equation}
 t_{ff}=\sqrt{\frac{3\pi}{32G\rho_{x}}}\approx{0.03 ~ {\rm s} \ \left(\frac{10^{9} \rm \frac{g}{cm^{3}}}{\rho_{wd}}\right)^{\frac{1}{2}} \left(\frac{r}{r_{th}}\right)^{\frac{3}{2}}} 
  \label{tff}
\end{equation}
Where $\rho_{x}$ is the density of the dark matter sphere which will be greater than the white dwarf density after collapse begins. The rightmost expression is obtained when $\rho_{x}=\rho_{wd}$, $i.e.$ when enough dark matter has been collected to achieve self-gravitation. 

The sound-crossing time, on the other hand, is given by the time a sound wave crosses the radius of the dark matter sphere. In a standard Jeans mass collapse analysis, if the sound-crossing time is longer than the free-fall time, a gravitationally bound configuration of gas will be unstable to small perturbations and will collapse. 

A self-thermalized dark matter sphere will have a short sound-crossing time. For our purposes, we will consider the dark matter sphere to be self-thermalized, since such a sphere will be most stabilized against collapse. If the dark matter is self-thermalized, it will have a sound speed given by, $c_{dm}\sim{\sqrt{T/m_{X}}}$. In such case,
\begin{equation}
 t_{sound}=\frac{r}{c_{dm}} \lesssim \frac{r_{th}}{c_{dm}}\approx{0.10 \ {\rm s} \ \left(\frac{ 10^{9} \rm{\frac{g}{cm^{3}}}}{\rho_{wd}}\right)^{\frac{1}{2}}},
 \label{eq:ts}
\end{equation}
By setting $r=r_{th}$, we have assumed that the dark matter sphere has just reached the point of self-gravitation, $i.e.$ that it has the same density as the white dwarf.  Comparing to Eq.~\eqref{tff}, the sound-crossing time for a self-thermalized dark matter sphere turns out to be longer than the free-fall time for a self-gravitating sphere of dark matter, for the white dwarf parameters considered in this paper. For lighter and correspondingly less dense white dwarfs and particular dark matter models, which will be treated in future work, it will be necessary to determine whether additional dark matter must be captured to satisfy the Jeans criterion. 

Before continuing, we briefly consider two relevant timescales: the time between dark matter self-interactions and the time between dark matter interactions with nuclei, prior to collapse. For much of the dark matter model space considered in this paper, the free-fall time in Eq.~\eqref{tff} is longer than the time it takes for a dark matter particle to interact with another dark matter particle, just prior to collapse,
\begin{align}
t_{XX} &= \frac{1}{n_X \sigma_{XX} \sqrt{T/m_X}} =  \frac{m_X}{\rho_{wd} \sigma_{XX} \sqrt{T/m_X}} \nonumber  \\&\approx 0.04~{\rm s}
~\left(\frac{10^{-30}~{\rm cm^2}}{\sigma_{XX}}\right)  \left(\frac{m_X}{10^6 \ {\rm GeV}}\right)^{\frac{3}{2}} \left(\frac{10^{9} \rm{\frac{g}{cm^{3}}}}{\rho_{wd}}\right)  \left(\frac{10^7~{\rm K}}{T}\right)^{\frac{1}{2}}
\end{align}
where $n_{X}=\rho_{wd}/m_{X}$ is the dark matter number density just before collapse and $\sigma_{XX}$ is the dark matter self-interaction cross-section. This shows that, in order for the dark matter to be self-thermalized on timescales faster than the collapse timescale, its self-interaction cross section must exceed $\sim 10^{-30}~{\rm cm^2}$. In Section \ref{sec:models} we will see that this condition is fulfilled for some explicit vector- and scalar-portal dark matter models.

At the onset of collapse, the time it takes for dark matter to interact with a nucleus in the white dwarf is
\begin{align}
t_{aX} &= \frac{1}{n_{a} \sigma_{aX} v_{nuc}} \approx  \frac{3 m_{a} T}{2 \rho_{wd} A^{4} E_{sup} \sigma_{nX} \sqrt{2E_{sup}/m_a}} \nonumber \\&\approx 10~{\rm s}
~ \left(\frac{10^{-40}~{\rm cm^2}}{\sigma_{nX}}\right)  \left(\frac{m_X}{10^6~ {\rm GeV}}\right)^{\frac{3}{2}} \left(\frac{10^7~{\rm K}}{T}\right)^{\frac{5}{2}}
\end{align}
where we conservatively set the Helm form factor to unity in $\sigma_{aX}$, and also evaluated the structure factor \eqref{eq:S''} at the thermal energy $3T/2$ of the dark matter. Comparing this to the free-fall time $t_{ff}$, we see that the above Jeans instability requirement was justified in neglecting dark matter-nuclear interactions.

The capture rate, thermalization rate, and conditions for dark matter collapse, all depend on the dark matter-nucleon cross section. Using the Montreal White Dwarf Database \cite{2017ASPC}, we find that most of the single white dwarfs in the mass range $1.2M_{\odot}<M<1.4M_{\odot}$ have cooling ages that lie below 3 Gyrs. The oldest $M\sim{1.4M_{\odot}}$ white dwarf present in the sample has an age of approximately ${3.3}$ Gyrs. Therefore, we will conservatively constrain the cross section by requiring both thermalization and accumulation of the critical mass to be completed within 3 Gyrs. Figure \ref{fig:therm} displays the minimum cross section in a log-log plot for all relevant processes to be completed for an $M\sim{1.4M_{\odot}}$ white dwarf of typical parameters within that time. 

\section{Collapsing dark matter energy transfer and ignition}
\label{sec:ignition}
Once the dark matter sphere begins to implode it will scatter with nuclei as it collapses, and can transfer enough energy to heat a small central region of the white dwarf to the point that nuclear fusion begins, resulting in a Type Ia supernova explosion \cite{Bramante:2015cua}. The major requirements for dark matter ignition via elastic scattering during dark matter collapse are as follows
\begin{enumerate}
\item The dark matter must heat the white dwarf faster than it cools (mostly via electron conduction cooling).
\item The dark matter must heat a portion of the white dwarf to a critical temperature (around $10^{10}$ K). 
\end{enumerate}
Below we will find that for dark matter elastically scattering with nucleons, satisfying the first criterion guarantees satisfying the second.

\subsection{Dark matter heating versus white dwarf cooling}
To begin, we note that Helm form factor coherence drops for higher energy scatters of dark matter against nuclei, whereas the structure factor \eqref{eq:S''} evaluates to unity. As the dark matter sphere collapses, the dark matter will move faster and faster, resulting in higher energy collisions. This allows us to determine a collapse radius past which dark matter energy transfer via elastic scattering with nuclei becomes suppressed (as we will see, the Helm form factor becomes relevant in this fashion for dark matter masses $m_X \gtrsim 10^8$ GeV). In future work, energy transfer via high energy dark matter-nucleus inelastic scattering may be considered; here we conservatively require that dark matter ignite white dwarfs during collapse purely through elastic scattering.

First, we calculate the velocity of the dark matter particles in the collapsing sphere, assuming a homogeneous density profile and applying the virial theorem
\begin{equation}
v_{vir}=\sqrt{\frac{3}{5}\frac{G N_{sg} m_{X}}{r}}\approx{6.8\times{10^{-7}} \left(\frac{10^{6} \ \gev}{m_{X}}\right)^{\frac{3}{4}} \left(\frac{10^{9} ~{\rm \frac{g}{cm^3}}}{\rho_{wd}}\right)^{\frac{1}{4}} \left(\frac{T}{10^7 \ \rm K}\right)^{\frac{3}{4}} \left(\frac{r_{th}}{r}\right)^{\frac{1}{2}}}
\end{equation}
The Helm form factor, on the other hand, begins to suppress energy exchanges at speeds $v \sim{0.02}$. Combining this with the previous expression, we obtain a radius at which the energy transferred per dark matter-nuclear collision begins decreasing,
\begin{equation}
 r_{min}\approx{ 9.5 \times 10^{-6} \ {\rm cm} \  \left(\frac{10^{6} \ \gev}{m_{X}}\right)^{\frac{3}{2}}\left(\frac{ 10^{9} ~{\rm \frac{g}{cm^3}}}{\rho_{wd}}\right)^{\frac{1}{2}}\left(\frac{T}{10^7 \ \rm K}\right)^{\frac{3}{2}}}.
\end{equation}
We will use this radius as a benchmark collapse position, for which the dark matter elastic scattering energy transfer to white dwarf nuclei is maximized.

To determine whether dark matter sparks nuclear fusion that leads to a Type Ia supernova, we follow the analysis of \cite{1992Timmes,Bramante:2015cua} and require the heating of a central mass of carbon and oxygen to a certain critical temperature $T_{c}$. Because the dark matter sphere encloses a small radius, we will mostly consider the heating of a microgram (1 $\mu$g) mass sample of white dwarf material composed of predominantly $^{12}$C nuclei. As determined in \cite{1992Timmes}, a microgram of white dwarf material must be heated to a temperature $T_{c}\sim{10^{10}}$ K ($\sim{1 \ \mev}$) in order to spark a supernova. This amount of white dwarf material will be enclosed in a radius
\begin{equation}
 r_{\mu{g}}\approx{7\times{10^{-6}} \ {\rm cm} \left(\frac{10^{9} ~{\rm \frac{g}{cm^3}}}{\rho_{wd}}\right)^{\frac{1}{3}}}.
\end{equation}
For dark matter heavier than $m_X \gtrsim 10^6$ GeV, we see that the radius of maximum coherent elastic scattering heating $r_{min}$ is smaller than the radius that encloses a microgram of white dwarf material; hence it will be convenient to simply consider how much the dark matter heats a microgram of white dwarf material. For $m_{X}\lesssim{10^{6} \ \gev}$, where $r_{\mu{g}}<{r_{min}}$, we have independently verified that that the cross-section required for dark matter to accumulate to a critical mass within $\sim{3}$ Gyrs, is sufficient to prompt Type Ia ignition, see also Ref.~\cite{Bramante:2015cua}. Therefore, we consider dark matter collapsing to radius $r_{min}$, which is smaller than the radius enclosing a microgram sample. For high dark matter masses, $m_X \gtrsim 10^8$ GeV, we will also see how increasing the dark matter-nucleon cross section beyond the cross section required to accumulate a critical mass of dark matter, allows for ignition during collapse. Compared to prior results in \cite{Bramante:2015cua}, this will improve the cross section bound obtained.

To determine the cross section required to ignite the sample, we analyse four relevant quantities: the energy transfer rate of the dark matter sphere to nuclei via elastic scattering $\dot{Q}_{DM}$, the diffusion rate at which the white dwarf sample cools down $\dot{Q}_{Diff}$, the heat capacity of the white dwarf sample $C_{v}$, and the latent heat absorbed $Q_{lat}$ when the white dwarf material melts from heating.

To calculate the energy transfer rate, we first consider the average energy exchange between a dark matter particle and a nucleus of mass $m_{a}$ in a single scatter. In the limit $m_{a}\ll{m_{X}}$ this energy exchange is simply $\Delta{E}\sim{m_{a} v_{vir}^{2}}$. Using the time it takes for a dark matter particle to interact with a nucleus $t_{aX}$, we write the energy transfer rate as
\begin{equation}
 \dot{Q}_{DM}=\frac{N_{sg} \ \Delta{E}}{t_{aX}}\approx{2 \times{10^{37}} \  { \rm \frac{GeV}{s}} \ \left(\frac{\sigma_{nX}}{10^{-40} \ \rm cm^{2}}\right) \left(\frac{10^{6} \ \gev}{m_{X}}\right)^{\frac{5}{2}}  \left(\frac{\rho_{wd}}{10^{9} ~{\rm \frac{g}{cm^3}}}\right)^{\frac{1}{2}} \left(\frac{T}{10^7 \ \rm K}\right)^{\frac{3}{2}}}
 \label{qdm}
\end{equation}
The rightmost expression has been evaluated when $v_{vir}\approx{0.02}$, $i.e.$ when the energy transfer rate is approximately at its maximum before becoming suppressed by the Helm form factor. At such point, the scattering rate is dominated by the dark matter velocity, so the dark matter-nucleus scattering time is also evaluated as $t_{aX}=(n_{a}\sigma_{aX}v_{vir})^{-1}$. The structure factor \eqref{eq:S''}, on the other hand, evaluates to unity at such high energy exchanges. We note that $\dot{Q}_{DM}$ decreases with $m_{X}$. For higher dark matter masses, fewer particles need to be captured for the dark matter sphere to self-gravitate and collapse, so the resulting collapsing dark matter sphere will have fewer particles scattering with nuclei.

Next, we consider the rate at which heat is diffused out of the sample. The conductive diffusion rate of a white dwarf material sphere of radius $r_s$ at temperature $T_s$ is given by \cite{Shapiro:1983du}
\begin{equation}
 \dot{Q}_{Diff}=\frac{4 \pi^{2} T_s^{3}(T_s-T_{wd}) \ r_s}{15 \kappa_{c} \rho_{wd}}\approx{1.6 \times{10^{25}} \ { \rm \frac{GeV}{s}} \ \left(\frac{\rho_{wd}}{10^{9} ~{\rm \frac{g}{cm^3}}}\right)^{\frac{4}{15}}}
 \label{diff}
\end{equation}
Where $\kappa_{c}\sim{10^{-9} \ {\rm cm^{2}/g} \ (T_{c}/10^{7} K)^{2.8} \ (10^{9} ~{\rm \frac{g}{cm^3}} \ /\rho_{wd})^{1.6}}$ is the conductive opacity of the white dwarf, which for densities above $\sim 10^7 {\rm g/cm^3}$ is dominated by relativistic electron conduction \cite{Potekhin:1999yv,Bramante:2015cua}. In \eqref{diff}, $T_{wd}\sim{10^{7}}$ K denotes the core equilibrium temperature. On the right hand side of Eq.~\eqref{diff}, we have set $T_s=T_{c}$ and $r_s=r_{\mu{g}}$ to indicate the scaling of the diffusion rate $\dot{Q}_{Diff}$, since diffusion will be maximized at the highest temperatures. Put another way, the choice $T_s=T_{c}$ yields the highest possible diffusion rate for the sample of white dwarf material prior to ignition.

In order to be ignited, the dark matter must heat the white dwarf faster than it can cool itself. The sample of white dwarf material can be ignited so long as the energy transfer rate (\ref{qdm}) exceeds the diffusion rate (\ref{diff}),
\begin{equation}
 \dot{Q}_{DM}>\dot{Q}_{Diff}.
  \label{ign}
\end{equation}
This condition ensures that the white dwarf material can be heated to $10^{10}~{\rm K}$, given the conductive electron diffusion in the white dwarf. For dark matter with a mass in the range $m_X \sim 10^8-3 \times 10^{12}$ GeV, this condition (Eq.~\eqref{ign}) places the most stringent restriction on the dark matter-nucleon cross section . This is because, for dark matter masses below $m_X \sim 10^{8}$ GeV, the cross section on nucleons required to accumulate a self-gravitating mass of dark matter is sufficient to prompt white dwarf ignition when the dark matter sphere collapses. On the other hand, for dark matter masses above $m_X \sim 3 \times 10^{12}$ GeV, ignition via black hole evaporation is guaranteed, as will be explained in Section \ref{sec:bhignition}. For dark matter in the mass range $m_X \sim 10^8-3 \times 10^{12}$ GeV, the ignition cross section increases with higher dark matter masses, since fewer dark matter particles will compose the collapsing dark matter sphere (see Eq. \eqref{eq:nsg}), and therefore, the overall rate for dark matter scattering against nuclei declines, resulting in a lower energy transfer rate for a fixed dark matter-nucleon scattering cross section. Consequently, the constraint on the cross section associated with ignition has a positive slope.  

We make the following remark about possible quantum mechanical effects: comparing the mean de Broglie wavelength of the dark matter particles $\lambda_{dB}=1/m_{X}v_{vir}$ to the mean dark matter particle separation $\sim{r/N_{sg}^{1/3}}$, we find that the dark matter sphere must collapse to a radius 
\begin{equation}
 r_{qm}\sim{\frac{1}{G m_{X}^{3} N_{sg}^{1/3}}}\approx{{10^{-6}} \ {\rm{cm}} \ \left(\frac{10^{6} \ \gev}{m_{X}}\right)^{\frac{13}{6}}\left(\frac{\rho_{wd}}{ 10^{9} ~{\rm \frac{g}{cm^3}}}\right)^{\frac{1}{6}}\left(\frac{10^7 \ \rm K}{T}\right)^{\frac{1}{2}}},
\end{equation}
in order for quantum mechanical effects to become relevant. This radius is much smaller than either $r_{min}$ or the Schwarzchild radius in the case of black hole evaporative ignition (see next section), for the dark matter masses considered in this paper. Therefore, we do not consider the effect of fermion or boson degeneracy in the dynamical collapse of dark matter inside the white dwarf.

\begin{figure}[h!] 
\includegraphics[scale=0.7]{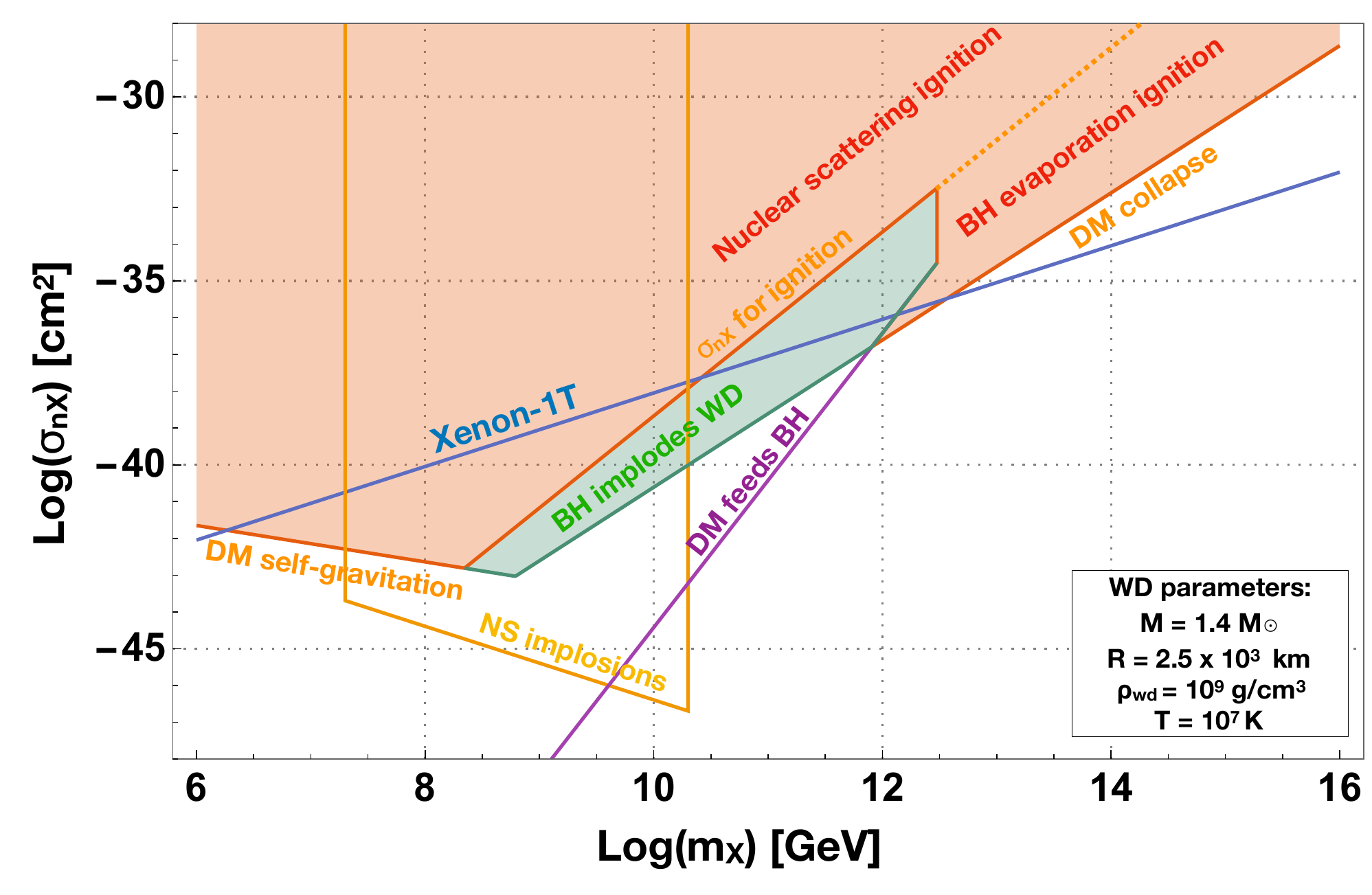} 
\caption{Constraints on the dark matter-nucleon cross section $\sigma_{nX}$, as a function of $m_{X}$, for Type Ia supernova ignition within 3 Gyrs, by either nuclear scattering or black hole evaporation (orange). Lines labelled ``DM self-gravitation" (orange/green) and ``$\sigma_{nX}$ for ignition" (orange), denote the minimum $\sigma_{nX}$ for white dwarf ignition by nuclear scattering, either from requiring that enough dark matter is captured so that it ``self-gravitates", or that collapsing dark matter scatters frequently enough to ignite the white dwarf. The line labelled ``DM collapse" (orange/green) limits the parameter space where dark matter will not ignite the white dwarf during collapse, but instead collapses and forms a black hole. Complete evaporation of this black hole via Hawking radiation ignites the white dwarf. The line ``DM feeds BH" (purple) denotes the limiting region where black hole evaporation is potentially stalled by further dark matter capture (see Figure \ref{fig:therm}). The region ``BH implodes WD" (green) is the parameter space where a black hole that consumes the white dwarf would be formed. A dark matter halo density $\rho_{X}\approx 0.3 \ \rm{GeV/cm^{3}}$ at the location of the white dwarf is assumed. All bounds were obtained using the parameters displayed in the lower right, taken from an old white dwarf (SDSS J160420.40+055542.3) selected to provide the most stringent bound for $m_X \lesssim 10^8$ GeV, out of all single white dwarfs listed in the Montreal White Dwarf Database \cite{2017ASPC}. See Appendix \ref{app:wds} for an extended discussion of bounds obtained from different white dwarfs, including those in old globular clusters \cite{McCullough:2010ai}. Also displayed are the bounds obtained from the Xenon-1T experiment (blue) \cite{Aprile:2017iyp} and neutron star implosions (yellow) \cite{Bramante:2017ulk}.}
\label{fig:bounds2}
\end{figure}

\subsection{Dark matter collapse and heating to critical temperature}
To determine the total amount that dark matter heats the white dwarf, we first determine the timescale for dark matter to collapse to $r_{min}$. We compute this by integrating the ratio between the rate at which the dark matter sphere sheds gravitational energy and the rate at which energy is transferred to the white dwarf. Collapse of the dark matter sphere will be again divided in two regimes, depending on whether the energy of the dark matter particles lies above or below the suppression scale $E_{sup}$. During the initial stages of collapse, dark matter is moving very slowly resulting in low momentum transfers and a high suppression of nuclear interactions, as we saw above. Dark matter exits this regime when it has collapsed to a radius

\begin{equation}
 r_{sup}=\frac{GN_{sg}m_{X}^{2}}{E_{sup}}\approx 10^{-3} \ {\rm cm} \left(\frac{10^{6} \ \gev}{m_{X}}\right)^{\frac{3}{2}} \left(\frac{10^{9} \ \rm{g/cm^{3}}}{\rho_{wd}}\right)^{\frac{3}{2}} \left(\frac{T}{10^{7} \ \rm{K}}\right)^{\frac{5}{2}},
\end{equation}

such that the nuclear scattering suppression is lost. We note that $r_{min}\ll r_{sup}\ll r_{th}$ always across the parameter space under consideration. Once the dark matter sphere collapses to $r_{sup}$, it will continue to collapse at a much faster rate. We find the collapse timescale to be determined by the time spent during this first regime, which we obtain by integrating Eq.~\eqref{eq:th-vis} from an initial energy $GN_{sg}m_{X}^{2}/r_{th}$ to an energy $GN_{sg}m_{X}^{2}/r_{sup}$,

\begin{equation}
 t_{col}\approx \frac{\sqrt{m_{a}}E_{sup}r_{th}}{2\sqrt{T}GN_{sg}m_{X}\rho_{wd}A^{4}\sigma_{nX}}\approx 10 \ \rm{yrs} \left(\frac{10^{-40} \ {\rm cm^{2}}}{\sigma_{nX}}\right)\left(\frac{m_{X}}{10^{6} \ \gev}\right)^{2}\left(\frac{10^{7} \ \rm K}{T}\right)^{\frac{5}{2}}
 \label{eq:collapse-time}
\end{equation}

In the rightmost expression, we have again set the Helm form factor to unity. The minimum cross section for the dark matter sphere to collapse within 3 Gyrs is identical to that required for completing the second thermalization phase, and is displayed in Figure \ref{fig:therm}.

The final requirement for ignition we now address, is that enough total energy should be transferred to the white dwarf material, so that it is heated to a temperature of $10^{10}~{\rm K}$ \cite{1992Timmes}. We will see that sufficient ignition heating for a microgram of white dwarf is guaranteed for the dark matter masses and cross sections we consider, so long as Eq.~\eqref{ign} is satisfied. The heat capacity of the sample is largely governed by the capacitance of the degenerate white dwarf's ion lattice \cite{Shapiro:1983du}
\begin{equation}
 C_{v}=\frac{3}{2}\left(\frac{4}{3}\pi{r_{\mu{g}}^{3}}n_{a}\right)\approx{746 \ \frac{\rm GeV}{\rm K}}
\end{equation}
Note that in our treatment the heat capacity does not scale with the core density, since we have fixed the sample mass to 1 $\mu$g. Then the total energy that has to be imparted to a microgram of white dwarf material is simply $C_{v}T_{c}\sim{10^{13} \ \gev}$. 

The critical temperature for ignition lies above the melting temperature $T_{m} \sim 8.5 \times 10^{8}$ K of the white dwarf material, so we must also consider the latent heat absorbed when the microgram material undergoes a solid-liquid phase transition. The amount of latent heat absorbed in the case of heating a microgram of white dwarf material is

\begin{equation}
 Q_{lat}\approx T_{m}\left(\frac{4}{3}\pi{r_{\mu{g}}^{3}}n_{a}\right)\approx 50 \ \gev
\end{equation}
We see that dark matter ignition of the white dwarf is not limited by the amount of potential energy the dark matter has available to be transferred: both $Q_{lat}$ and $C_{v}T_{c}$ are much smaller than either $\dot{Q}_{DM}$ or $\dot{Q}_{Diff}$ multiplied by the collapse timescale over which dark matter potential energy is transferred.

We make one final comment on the ignition of Type Ia supernovae by dark matter. It has been noted that many deflagration and detonation models of ignition \cite{Niemeyer:1996pm}, require that the heated white dwarf material to have a temperature profile with a high degree of homogeneity. The time scale for energy diffusion is
\begin{equation}
 t_{Diff}=\frac{\kappa_{c}\rho}{r_{\mu{g}}T_{c}^{3}}\approx{1.4\times{10^{-28}} ~{\rm{s}} \ \left(\frac{10^{9} ~{\rm \frac{g}{cm^3}}}{\rho_{wd}}\right)^{2.27}}
 \label{tdiff}
\end{equation}
We note that $t_{Diff}$ is shorter than the time scale for energy transfer to the white dwarf material. This shows that energy diffusion within the microgram sample is sufficiently fast to ensure an homogeneous temperature profile of the sample during the heating process. 

The final constraint on $\sigma_{nX}$, inferred from the existence of an old white dwarf that has not exploded, is plotted on Figure \ref{fig:bounds2} for all the mass range considered throughout this work. A sizable region below the Xenon-1T results \cite{Aprile:2017iyp} is excluded, which has some overlap with bounds derived from neutron star implosions \cite{Bramante:2017ulk}.  

\section{Black hole formation and evaporative ignition}
\label{sec:bhignition}
We now consider the possibility that the dark matter sphere collapses to form a black hole, which then ignites the white dwarf while evaporating via Hawking radiation. For a fixed dark matter-nucleon cross-section this will occur for heavier dark matter, since the number of dark matter particles and the power transferred to the white dwarf both decrease as the dark matter mass $m_{X}$ increases, $cf.$ Eq.~\eqref{qdm}. We also analyse the parameter space region in which the black hole formed does not evaporate, but rather grows by accretion of both dark matter and stellar material, eventually consuming the white dwarf.

\subsection{Black hole formation from collapsing dark matter}
The mass of dark matter required to form a black hole will depend on the spin of the dark matter field and its self-couplings \cite{Colpi:1986ye,Kouvaris:2015rea,Gresham:2018rqo}. In the case of a Higgs portal dark matter model examined in Section \ref{sec:models}, the dark matter self-interactions are attractive and do not preclude the black hole formation. Self-interactions via a vector boson present a more complex case, as they can be either attractive or repulsive. In the absence of a detailed study of these interactions, we conservatively use the maximum mass stabilized against non-relativistic gravitational collapse by Fermi-degeneracy pressure for dark matter fermions
\begin{align}
M_{BH,min}^{ferm} = \frac{M_{pl}^3}{m_X^2} \approx{2\times{10^{37} \ \gev} \left(\frac{10^{10} \ \gev}{m_{X}}\right)^{2}}
\end{align}
and the maximum mass stabilized against non-relativistic gravitational collapse by quantum pressure for bosons with an order unity quartic self-coupling ($\lambda_\phi \sim 1$ for $V(\phi) \supset \lambda_\phi \phi^4$) \cite{Colpi:1986ye}
\begin{align}
M_{BH,min}^{boson} = \frac{\sqrt{\lambda_\phi} M_{pl}^3}{m_X^2} \approx 2 \times 10^{37}\ \gev\left(\frac{10^{10} ~{\rm GeV}}{m_X} \right)^{2}  \left(\frac{\lambda_\phi}{1} \right)^{1/2}.
\end{align}
Comparing these minimum black hole masses to the mass required for dark matter to self-gravitate obtained from Eq.~\eqref{eq:nsg},
\begin{align}
M_{crit}\approx 6.3\times{10^{37}~{\rm GeV}} \left(\frac{10^{10} \ \gev}{m_{X}}\right)^{\frac{3}{2}} \left(\frac{10^{9} {\rm \frac{g}{cm^3}}}{\rho_{wd}}\right)^{\frac{1}{2}} \left(\frac{T}{10^7{\rm  K}}\right)^{\frac{3}{2}}.
\label{eq:mcrit}
\end{align}
we find that if dark matter heavier than $10^{10}$ GeV collapses under its own weight at the center of a white dwarf with the parameters specified, it will be massive enough to form a black hole.

The critical mass for dark matter collapse can be used to compute the final black hole's Schwarzchild radius
\begin{equation}
 r_{s}=2GM_{crit}\approx{1.3\times{10^{-14} \ {\rm cm}} \ \left(\frac{10^{10} \ \gev}{m_{X}}\right)^{\frac{3}{2}} \left(\frac{10^{9} ~{\rm \frac{g}{cm^3}}}{\rho_{wd}}\right)^{\frac{1}{2}}\left(\frac{T}{10^7 K}\right)^{\frac{3}{2}}}
\end{equation}
We remark that the Schwarzschild radius is much smaller than the radius of the white dwarf enclosing a microgram of material, $r_{s}\ll r_{\mu{g}}$ for all $m_{X}>10^{10} \ \gev$, which will be relevant for calculating white dwarf ignition via black hole evaporation. 

The time required for the black hole to be formed is obtained from Eq.~\eqref{eq:collapse-time}. This time imposes the most stringent constraint on $\sigma_{nX}$, when setting a bound on dark matter parameters using old white dwarfs. The minimum cross section for dark matter collapsing and forming a black hole within 3 Gyrs is displayed in Figure \ref{fig:therm}. Below we will see, for dark matter masses from $m_X \sim 10^{12} - 10^{16}$ GeV, all black holes formed from dark matter with $r_{s} \lesssim 10^{-14} ~{\rm cm}$ will ignite $M \lesssim 1.4M_{\odot}$ white dwarfs as they evaporate.

\subsection{Black hole evaporative ignition of white dwarfs}
After formation, the growth and evaporation of the black hole is governed by
\begin{equation}
 \dot{M}_{bh}=\frac{4\pi\rho_{wd}{(GM_{bh})^{2}}}{c_{wd}^{3}}-\frac{1}{15360\pi{(GM_{bh})^{2}}} +f_X m_X C_X.
  \label{eq:bhtotal}
\end{equation} 
The first term on the right hand side is the Bondi accretion rate, where $c_{wd}\approx{0.03}$ is the sound speed at the core of the white dwarf, which has been computed using a polytropic equation of state \cite{Shapiro:1983du}. The second term accounts for mass loss through the emission of Hawking radiation. The third term governs the growth of the black hole from the infall of additional dark matter inside the white dwarf \cite{Bramante:2013hn}. Dark matter masses $m_{X}\gtrsim 10^{10} \ \gev$ form black holes formed in white dwarfs with initial masses $M_{crit}=N_{sg}m_{X}$ such that the Bondi accretion rate is negligible compared to the Hawking term. Therefore, for the parameter space of interest, the dark matter capture rate compared to the Hawking radiation rate will determine whether the black hole evaporates or grows. 

The dark matter capture rate on a small black hole at the center of the white dwarf will depend on a number of factors. If the dark matter is self-thermalized (see Section \ref{sec:capture}), the capture rate can be obtained by calculating the sound speed of the dark matter and determining its Bondi accretion rate on the black hole. If the dark matter cannot be modelled as an ideal gas, simulations or analytic estimates \cite{Doran:2005vm} of the capture rate using individual particle trajectories would be necessary. Here we will conservatively assume that the dark matter efficiently feeds the small black hole, $i.e.$ that $f_X =1$ in Eq.~\eqref{eq:bhtotal}. This will be a conservative assumption for obtaining bounds using black hole evaporation ignition, because dark matter feeding the black hole can prevent the black hole from evaporating, which reduces its temperature and power output, thus inhibiting black hole evaporative ignition of the white dwarf. 

The time for a black hole with an initial mass $M_{crit}$ ($i.e.$ the self-gravitating mass) to evaporate, when neglecting the Bondi accretion rate and setting $f_{X}=1$ in Eq.~\eqref{eq:bhtotal}, is
\begin{equation}
 t_{bh}=\frac{M_{crit}}{m_{X}C_{X}}-\frac{a^{2} {\rm ArcTanh} \left[\sqrt{m_{X}C_{X}}M_{crit}/a^{2}\right]}{(m_{X}C_{X})^{3/2}},
 \label{eq:gensol}
\end{equation}
where we define a Hawking evaporation constant $a^{2}=(G\sqrt{15360\pi})^{-1}$. It can be shown that $\sqrt{m_{X}C_{X}}M_{crit}<a^{2}$ implies a rapid black hole evaporation time. In this limit, $\sqrt{m_{X}C_{X}}M_{crit}\ll{a^{2}}$, Equation \eqref{eq:gensol} approximates to 
\begin{equation}
 t_{bh}=15360\pi  G^{2} M_{crit}^{3}\approx{10^{9} \ {\rm yrs} \ \left(\frac{1.5 \times 10^{10} \ \gev}{m_{X}}\right)^{\frac{9}{2}} \left(\frac{10^{9} ~{\rm \frac{g}{cm^3}}}{\rho_{wd}}\right)^{\frac{3}{2}} \left(\frac{T}{10^7 K}\right)^{\frac{9}{2}}},
 \label{bh3}
\end{equation}
This is the usual solution obtained when the Hawking radiation term dominates alone. On the righthand side of this expression we have used the self-gravitating dark matter mass \eqref{eq:mcrit} to determine how quickly the black hole formed in a white dwarf star evaporates. In practice, the solution \eqref{eq:gensol} converges very quickly to \eqref{bh3} in the limit mentioned above. The condition we require therefore is that black hole evaporation exceed dark matter capture onto the black hole,
\begin{align}
\frac{1}{15360\pi{(GM_{crit})^{2}}} > m_X C_X ,
\label{eq:bh-feeding}
\end{align}

This requirement is shown in Figure \ref{fig:bounds2}. For masses $m_{X}\gtrsim{10^{11} \ \gev}$, the capture of dark matter cannot prevent black hole evaporation for any given cross section. This is because the Hawking evaporation rate lies above the saturation value for the capture rate, $cf.$ Eq.~\eqref{eq:capturedm-approx}, so \eqref{eq:bh-feeding} is always fulfilled. Therefore, black holes formed from dark matter with mass $m_{X}\gtrsim 10^{11} \ \gev$ will inevitably evaporate. 

Of course, dark matter of with a mass smaller than  $m_{X} \sim 10^{12} \ \gev$ can also form evaporating black holes. We see from Eq.~\eqref{bh3} that black hole evaporation will occur in less than a billion years for $m_X \gtrsim 1.5 \times 10^{10}$ GeV, so long as the dark matter-nucleus cross section is small enough to satisfy Eq.~\eqref{eq:bh-feeding}. Therefore, we adopt $m_X=1.5\times 10^{10}$ GeV as a lower bound on the dark matter mass that can cause white dwarf evaporative black hole ignition for the white dwarf parameters given in Figure \ref{fig:bounds2}.

We now analyse black hole evaporative heating of a microgram-sized sample of white dwarf material, and determine when the conditions for ignition are met. First, the energy transfer rate of the black hole must exceed the rate at which heat diffuses out of the sample. So long as dark matter capture on the black hole is sufficiently small, the mass evolution will be completely determined by Hawking radiation emission. The Hawking radiation rate will increase as the black hole shrinks in size
\begin{align}
 \dot{M}_{bh} \approx 10^{30}~\frac{{\rm GeV}}{{\rm s}}~\left( \frac{10^{33}~{\rm GeV}}{M_{bh}}   \right)^2
 \label{eq:hawkrate}
\end{align}
Here we have normalized to a black hole mass $M_{bh} \sim 10^{33}$ GeV, which corresponds to an initial black hole mass for dark matter mass $m_X \sim 10^{13}$ GeV, $cf.$ Eq.~\eqref{eq:mcrit}. For black hole evaporation bounds derived in the paper, we require all formed black holes to evaporate completely. Therefore, for dark matter masses less than and greater than $m_X \sim 10^{13}$ GeV, this black hole evaporation heating rate will be matched and surpassed since we set bounds only on parameter space where the black holes evaporate completely, and the initial black hole mass shrinks with increasing dark matter mass (again see Eq.~\eqref{eq:mcrit}). Comparing Eq.~\eqref{eq:hawkrate} to the heat diffusion rate in a white dwarf, Eq.~\eqref{diff}, we see that the first requirement for black hole evaporative white dwarf ignition is satisfied, so long as a sizable portion of this energy is deposited in the white dwarf.

The second condition for evaporative black hole ignition of the white dwarf, is that the radiation emitted must effectively heat a sample of white dwarf materials to $\sim 10^{10}$ K. This will depend on the amount of energy deposited by Hawking emitted particles as they propagate across the microgram-containing region of white dwarf material. We find that this condition is satisfied for all black holes evaporating inside the white dwarf, so long as the mass of the black hole is large enough, $M_{bh} \gtrsim 10^{16}$ GeV. This condition is satisfied for all the dark matter parameters we consider. 

The temperature of the black hole depends on its mass $M_{bh}$. As a black hole evaporates, its temperature and radiated power increase according to
\begin{equation}
T_{bh}=\frac{1}{8\pi{GM_{bh}}}\approx{6325 \ \gev \ \left(\frac{m_{X}}{10^{13} \ \gev}\right)^{\frac{3}{2}} \left(\frac{\rho_{wd}}{10^{9} ~{\rm \frac{g}{cm^3}}}\right)^{\frac{1}{2}}\left(\frac{10^7 \ \rm{K}}{T}\right)^{\frac{3}{2}}}.
\label{bh2}
\end{equation}
After it is formed from collapsing dark matter in the white dwarf, it is evident that the black hole's temperature will increase as it evaporates. In the rightmost expression we made the replacement $M_{bh}=M_{crit}$, $i.e.$ we take the black hole initial mass to be the critical mass for self-gravitation of the captured dark matter. The energy and composition of the radiated particles will depend on this temperature \cite{MacGibbon:1990zk,MacGibbon:1991tj}. While a detailed simulation of black hole emission propagation through white dwarf material is beyond the scope of this work, we will find that emitted color-charged particles alone are enough to ignite the white dwarf ($i.e.$ emission of gluons and quarks which become mesons and hadrons during propagation). We will neglect contributions from, $e.g.$, emitted neutrinos and leptons, as these will not alter the results of our analysis. Our conclusions in this regard match the findings of Reference~\cite{Graham:2018efk}, which presented a detailed study of Standard Model particle showers produced in the decay and annihilation of very heavy dark matter particles in white dwarfs. 

At the temperatures relevant for ignition \eqref{bh2}, all the particles present in the Standard Model are emitted relativistically. Each type of particle (where each helicity state counts as a separate particle) is emitted roughly equally up to percent level corrections \cite{Page:1977um,Autzen:2014tza}, so long as each particle's mass is less than the black hole temperature. For the temperature indicated in Eq.~\eqref{bh2}, we will only need to consider emitted particles that carry color charge. At these temperatures, more than half the black hole evaporation power is emitted in relativistic color-charged particles \cite{Page:1977um,Autzen:2014tza}. 

Black hole radiated color-charged particles with energies $\gtrsim 10^3$ GeV, have a kinetic energy above carbon's nuclear binding energy $E_{bind}\sim{10 \ \mev}$. To analyze the interactions of these color-charged particles (quarks, gluons which create hadrons, mesons) with carbon nuclei in the white dwarf, we follow the standard procedure for estimating hadronic showering in materials \cite{Grieder:1339647}. We take the inelastic carbon scattering cross-section for the high energy particles to be the geometric cross-section of the carbon nucleus, $\sigma_{C} \sim 10^{-25}~{\rm cm^2}$. Each time a high-energy particle hits a carbon nucleus over a mean free path $l_{had} \sim (n_C \sigma_C)^{-1} \sim 10^{-7}~{\rm cm}$, it fractures the carbon nucleus and creates $\sim 2-3 \sim e$ high energy hadrons/mesons ($e$ here is Euler's number). Each of these inelastic by-products in turn strikes a carbon nucleus, and after $N_i$ such interactions, the resulting showered particles each have final energies $E_f$ of the order
\begin{align}
E_f = \frac{E_i}{e^{N_i}},
\end{align} 
where $E_i$ is the initial particle energy and $N_i \sim n_C \sigma_C l_{sh}$ is the number of inelastic scattering interactions the hadronic shower has undergone over a length $l_{sh}$ with a carbon number density $n_C$. The total mean free path of the emitted hadronic shower will be 
\begin{align}
 l_{sh} &=\frac{1}{n_{C}\sigma_{C}} ~{\rm Ln\left(\frac{E_i}{E_f}\right)} \nonumber \\ & \approx{3 \times 10^{-6} \rm ~ cm} ~ \left(\frac{5 \times 10^{31}/{\rm cm^3}}{n_C}\right)  \left(\frac{10^{-25}~{\rm cm^2}}{\sigma_C}\right)  {\rm Ln} \left[ \left(\frac{E_i}{10^3~{\rm GeV}}\right) \left( \frac{0.001~{\rm GeV}}{E_f}\right)     \right],
\end{align}
where we have normalized the carbon density to a typical white dwarf core density of $10^{9} ~{\rm g/cm^3}$. At each showering stage, the emitted hadrons scatter inelastically against nuclei, the resulting hadrons produced in the collision split the initial energy. At the relevant ignition energies ($\sim {\rm MeV}$), the majority of the final showered by-product particles are pions. Note that the length of the shower depends only logarithmically on the energy $E_{in}$ of the initial color-charged particle.

We see that the mean free path for a high-energy color-charged particle is much less than the radius of white dwarf enclosing a microgram of material $l_{had}\lesssim {r_{\mu{g}}} \approx 7 \times 10^{-6} ~{\rm cm}$, while the showering length is  $l_{sh}\lesssim{r_{\mu{g}}} \approx 7 \times 10^{-6} ~{\rm cm}$. Note that these statements hold even for the smallest black holes formed by dark matter in this study: for $m_X \sim 10^{17}$ GeV, $M_{bh} \sim 10^{27}~{\rm GeV}$, and $l_{sh} \sim 6 \times 10^{-6} ~{\rm cm}$. This implies that the color-charged particles emitted from the black hole travel a short distance, compared to the sample size, before scattering inelastically against nuclei. The showers produced have a somewhat larger energy deposition length, heating a $\sim \mu$g white dwarf region to MeV temperatures roughly uniformly during subsequent showering. Therefore, we find that color-charged particles emitted from the black hole constitute an efficient, isothermal heating mechanism that will prompt white dwarf ignition. We remark that, although the highest energy color-charged Hawking particles induce fission in nuclei upon collision, the final output particles in the hadronic shower (which exponentially outnumber the higher energy input particles) lack the energy to produce further fission. Therefore, we can safely assume that the critical temperature $T_{c}$ for ignition remains unchanged, as the majority of the $\sim \mu$g mass white dwarf sample preserves its composition as it heats up. 

Finally, while they are not necessary for ignition, we briefly comment on the emission of non-color-charged particles, and photons. The emitted photons scatter inelastically against nuclei, producing hadronic showers. At these energies, this is the dominant mechanism involved. The inelastic scatters generate these photonuclear showers after a somewhat longer mean free path, since they couple to nuclei through the electromagnetic rather than the strong force. Similarly, emitted electrons scatter inelastically via the exchange of a virtual photon. The difference compared to the previous cases discussed, is that the electron survives the interaction. Both electronuclear and photonuclear showers will contribute less to white dwarf heating than the black hole's hadronic emission, since for the black hole temperatures we consider, there is more emission of strongly-coupled particles.

We have now established that black holes with masses  $M_{bh} \sim 10^{27}-10^{33}~{\rm GeV}$, formed from collapsing dark matter in white dwarfs, will quickly and uniformly heat a microgram of white dwarf material to ignition via Hawking evaporation. It remains to point out that the large masses of these black holes imply that the energy they impart to the microgram mass vastly exceeds the $\sim 10^{13} \ \gev$ required for ignition, as discussed in Section \ref{sec:ignition}. 
Thus, we find that the white dwarf is ignited by dark matter collapsing and forming a black hole in its interior, for dark matter masses $m_{X}\gtrsim {1.5 \times \ 10^{10} \ \gev}$, so long as the black hole evaporates, which in turn depends on how much the black hole is fed by accreted dark matter. For this parameter space, the most restrictive constraint on the cross section comes from the requirement that the dark matter collapses and forms a black hole in less than a few gigayears. 

\subsection{Black hole destroying the white dwarf}

In addition to the previous analysis, there is a region of the parameter space where the black hole, fed by dark matter and white dwarf material, grows over time to consume the white dwarf star. Comparing the Bondi accretion rate to the Hawking evaporation rate, we see that a newly-formed black hole needs a minimum mass
\begin{equation}
M_{bondi}=\left(\frac{c_{wd}^{3}}{61440\pi^{2}\rho_{wd} G^{4}}\right)^{1/4}\approx 5\times 10^{37} \ \gev \left(\frac{10^{9} ~{\rm \frac{g}{cm^3}}}{\rho_{wd}}\right)^{\frac{1}{4}},
\label{eq:bondimass}
\end{equation}
for the accretion of stellar material to exceed the evaporation rate. The initial mass of the black hole will be larger than this mass for dark matter masses $m_{X}\lesssim 10^{10} \ \gev$, $cf.$ Eq.~\eqref{eq:mcrit}. In that case, for cross sections below those necessary for ignition via nuclear scattering, any black hole formed will accrete both dark matter and stellar material, quickly consuming the star in a short timescale. For heavier dark matter masses $ 10^{10} \ \gev \lesssim m_{X} \lesssim 10^{12} \ \gev$, the initial black hole mass is small enough that the Hawking evaporation rate dominates over accretion of stellar material. However, if the dark matter-nuclear cross section is large enough such that the black hole is fed efficiently, $cf.$ Eq.~\eqref{eq:bh-feeding}, captured dark matter feeds the black hole preventing it from evaporating. We find for all green-shaded parameter space in Figure \ref{fig:bounds2} that the black hole reaches a mass \eqref{eq:bondimass} in less than a few billion years before consuming the white dwarf. 

To summarize: if the dark matter-nucleon cross section is below that necessary for ignition via nuclear scattering, but at the same time large enough to prevent black hole evaporation, dark matter of mass $m_{X}\lesssim 10^{12} \ \gev$ will collapse forming a black hole that quickly destroys the star. In particular, the capture rate not only determines the time it takes to form a self-gravitating dark matter sphere, but also the rate at which the black hole is fed, preventing evaporation. Using an old white dwarf that has not exploded, we set bounds on the dark matter-nucleon cross section by requiring a self-gravitating dark matter sphere to be formed and collapse in no less than 3 Gyrs, as well as requiring that the black hole so formed destroys the white dwarf either through dark matter accretion, Bondi accretion of baryons, or both. This yields the additional green exclusion zone plotted in Figure \ref{fig:bounds2}.

\section{Dark matter models for ignition of white dwarfs}
\label{sec:models}
Having already treated dark matter's effect on white dwarfs using a generic per-nucleon cross section, we now investigate some explicit models of dark matter that cause white dwarfs to explode or implode. With late-decaying fields that deplete the late-time abundance of asymmetric dark matter, super-heavy asymmetric dark matter cosmologies have been constructed in \cite{Bramante:2017obj}. In the following models, we will consider light mediators and $\mathcal{O}(1)$ gauge coupling constants, implying sizable self-annihilation cross-sections for our candidate dark matter models. Using various methods that deplete late-time relic particle abundances, it would be possible to construct a cosmology for the dark matter models considered here, though precise model parameters satisfying these cosmologies will depend on a number of factors, for example the late-time energy density of decaying fields -- we leave a detailed treatment to future work. 

The first explicit dark matter model is a Dirac fermion $X$ as dark matter, which is charged under a new $U(1)_D$ gauge group with gauge coupling constant $\alpha_X$. The vector mediator $V_{\mu}$ of this gauge group mixes with the Standard Model photon, see $e.g.$ \cite{Holdom:1985ag,Pospelov:2007mp,Kaplinghat:2013yxa}. For this mixed-vector mediator model, often called ``hidden photon" or ``vector portal" dark matter, the coupling to the Standard Model is mediated through the following mixing terms in the Lagrangian 
\begin{equation}
 \mathcal{L}_{mix}=\frac{\epsilon_\gamma}{2}V_{\mu{\nu}}F^{\mu{\nu}}+\kappa^{2}V_{\mu}Z^{\mu}.
\end{equation} 
Here $V_{\mu{\nu}}=\partial_{\mu}V_{\nu}-\partial_{\nu}V_{\mu}$ is the $U(1)_{D}$ field-strength tensor, the first term is the dark-visible photon kinetic mixing term with mixing coupling $\epsilon_{\gamma}$, while the second term gives the mass mixing between the $U(1)_D$ vector boson and the Standard Model $Z$ boson with dimensionful coupling parameter $\kappa$.
The dark matter-nucleon scattering cross section against protons in this case is \cite{Kaplinghat:2013yxa}
\begin{equation}
 \sigma_{nX}^{p}\approx 2 \times 10^{-39} \ {\rm{cm^{2}}} \ \ \left(\frac{\alpha_{X}}{10^{-1}}\right) \ \left(\frac{\gev}{m_{\gamma}}\right)^{4} \ \left( \frac{\epsilon_\gamma}{10^{-5}} \right)^2
\end{equation}

\begin{figure}[t!] 
\includegraphics[scale=0.7]{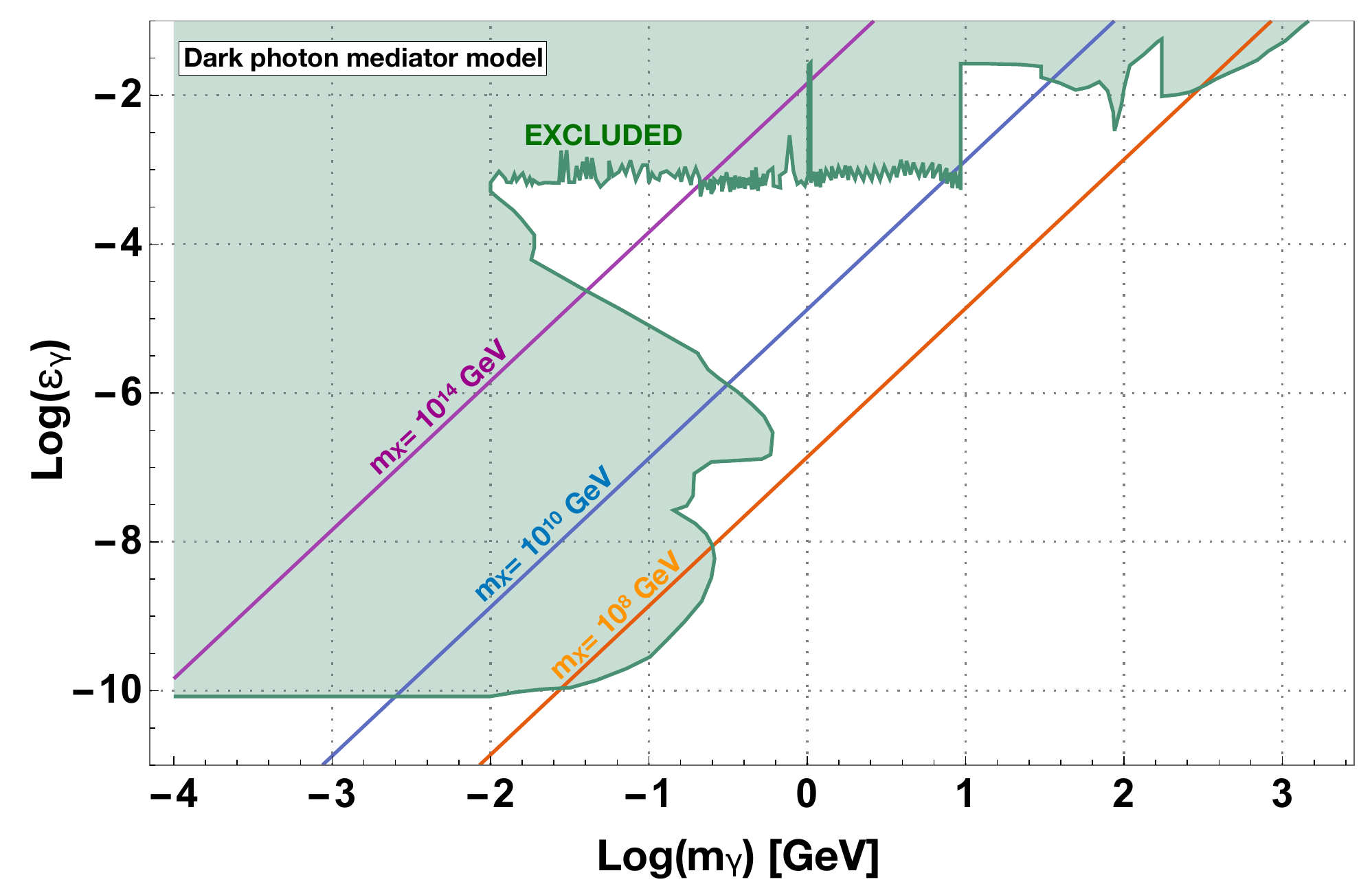} 
\caption{Constraints on the mixing parameter $\epsilon_{\gamma}$ for vector $\phi$ as a function of its mass $m_{\phi}$, in a log-log plot. We fix $\alpha_{X}=10^{-1}$ (note the linear dependence of the cross section on this parameter). The different colors correspond to different dark matter masses, as labeled. Regions above the solid lines are excluded by observations of the old white dwarf SDSS J160420.40+055542.3. Shaded regions denote parameter space bounded by supernova cooling and particle colliders \cite{Bramante:2016rdh,Abrahamyan:2011gv,Essig:2013lka,Goudzovski:2014rwa,Merkel:2014avp,Batell:2014mga,Curtin:2014cca}. We have omitted constraints on dark photons that would decay during, and thereby alter primordial nuclear abundances from big bang nucleosynthesis \cite{Fradette:2014sza}.}
\label{explmodv}
\end{figure}

This dark photon model presents a much faster thermalization time as it allows dark matter to lose energy through repeated scatterings against electrons in the white dwarf. The energy loss rate for scattering against unbound electrons is \cite{Bhoonah:2018gjb},

\begin{equation}
 \left(\frac{dE}{dt}\right)_{e^{-}}=\left(\frac{8\pi\alpha_{X}\alpha_{em}\epsilon_{\gamma}^{2}m_{e}^{2}\Delta{E}n_{e}v_{rel}}{(m_{\gamma}+q^{2})^{2}}\right){\rm{Min}}\left[\frac{q}{p_{F}},1\right]
 \label{eq:electrons}
\end{equation}
We note that although Eq.~\eqref{eq:electrons} is valid for unbound electrons, this can be reasonably applied to white dwarfs since electrons in the interior can be treated as a nearly free gas \cite{Shapiro:1983du}. We have also added a Pauli blocking factor ${\rm{Min}}[q/p_{F},1]$ accounting for the fact that electrons are degenerate within the white dwarf medium, with a Fermi momentum $p_{F}\sim(3\pi^{2}n_{e})^{1/3}=(18\pi^{2}n_{a})^{1/3}\sim 5 \ \mev$. This energy loss mechanism is far more efficient than nuclear scattering when the latter is suppressed by the structure factor, leading to much shorter thermalization and collapse times. For instance, the final phase of thermalization (where nuclear scattering is suppressed, $cf.$ Eq.~\eqref{eq:th-vis}) would be completed through electron scattering in a time
\begin{equation}
 t_{e^{-}}^{th} \approx \frac{p_{F}E_{sup}}{2 \pi \alpha_{X} \alpha_{em} \epsilon_{\gamma}^{2}n_{e}} \approx 10^{-5} \ {\rm{s}} \left(\frac{m_{X}}{10^{6} \ \gev}\right) \left(\frac{\rho_{wd}}{10^{9} ~{\rm \frac{g}{cm^3}}}\right)^{\frac{1}{3}} \left(\frac{10^7 \ \rm K}{T}\right)
\end{equation}
In the rightmost expression, we conservatively chose $\alpha_{X}\approx 0.1$, a dark photon mass $m_{\gamma}\approx 10^{-12}~{\rm eV}$ and mixing parameter $\epsilon_{\gamma}\approx 10^{-7}$. Note that this implies a much smaller dark photon mass than is shown in Figure \ref{explmodv}. Consequently, we see that a vector portal model coupled through a very light dark photon can be constrained using old white dwarfs that have not exploded, given the efficient transfer of dark matter kinetic energy via electron scattering, for a sufficiently low mass dark photon.

Next, we consider a dark matter model, which couples to the Standard Model via a Yukawa interaction with a scalar mediator $\phi$ that in turn mixes with the Higgs boson \cite{Burgess:2000yq}. Again we take the dark matter to be a Dirac fermion $X$. For this ``Higgs portal" dark matter model, the real scalar field $\phi$ mixes with the Higgs boson $H$ through terms of the form
\begin{equation}
 \mathcal{L}_{\phi}= - \alpha_\phi \phi X \bar{X} -(a\phi+b\phi^{2})\left|H\right|^{2}
\end{equation}
Where $H$ is the Standard Model Higgs doublet and $a,b$ are coupling constants, where $a$ has dimensions of mass and $b$ is dimensionless. In the limit $a,m_{\phi}\ll{v,m_{h}}$ where $v\approx{246 \ \gev}$ is the Higgs vacuum expectation value after electroweak symmetry breaking, the mixing angle is $\epsilon_{h}\approx{av/m_{h}^{2}}$, with the following term describing the resulting coupling of $\phi$ to Standard Model fermions
\begin{equation}
 \mathcal{L}_{int}=-\frac{m_{f}}{v}f\bar{f}\phi
\end{equation}

\begin{figure}[t!]
\includegraphics[scale=0.7]{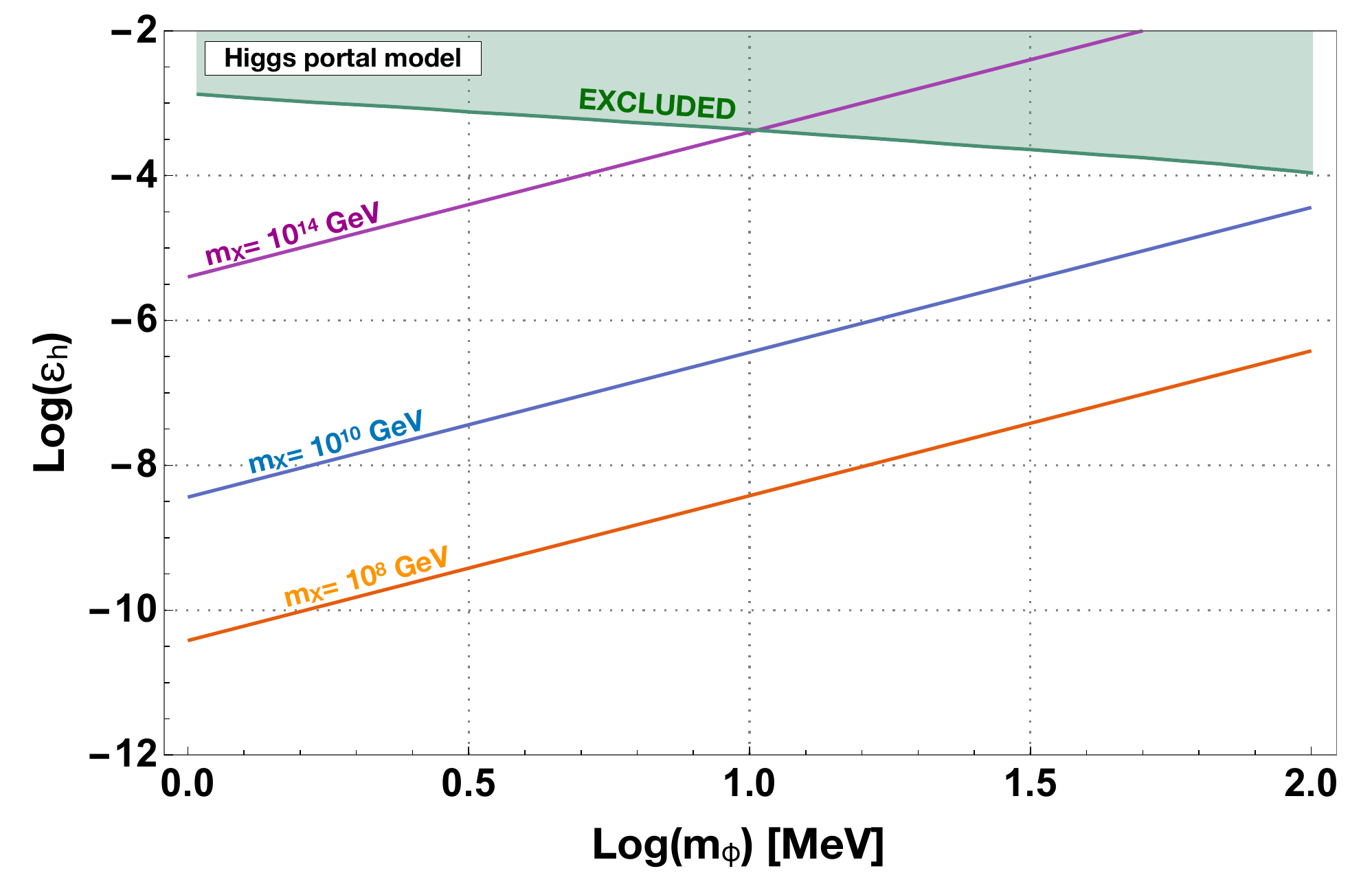} 
\caption{Constraints on the mixing parameter $\epsilon_{h}$ for scalar $\phi$ as a function of its mass $m_{\phi}$, in a log-log plot. We fix $\alpha_{\phi}=10^{-1}$ (note the linear dependence of the cross section on this parameter). The different colors correspond to different dark matter masses; regions above these coloured lines are excluded by observation of the old white dwarf SDSS J160420.40+055542.3. Shaded parameter space is excluded by non-observation of anomalous meson decays at the CHARM experiment \cite{Dolan:2014ska}. There are additional constraints on the Higgs portal scalar from cosmological considerations, including contributions to relativistic degrees of freedom, distortions of the cosmic microwave background, and primordial nucleosynthesis abundances \cite{Fradette:2018hhl}.}
\label{explmods}
\end{figure}

Both of the above dark sector mediators, and their respective mixing mechanisms allow $V_{\mu}$ and $\phi$ to decay in the early Universe, and constraints on the mixing parameters can be obtained from requiring Big-Bang nucleosynthesis to be unaffected by these decays and the resulting dilution of baryon density \cite{Kaplinghat:2013yxa}. Additional constraints for the mixing parameters in the vector case are obtained from meson decay experiments, as shown in Figure \ref{explmods}. 

For the case of a scalar mixed with the Higgs boson, the nucleon scattering cross-section is
\begin{equation}
 \sigma_{nX}\approx 2 \times 10^{-38} \ {\rm{cm^{2}}} \ \ \left(\frac{\alpha_{\phi}}{10^{-1}}\right) \ \left(\frac{30 \ \mev}{m_{\phi}}\right)^{4} \  \left(  \frac{\epsilon_{h}}{10^{-5}} \right)^{2}
\end{equation}
Using the white dwarf explosion bounds obtained for the dark matter-nucleon cross section in Section \ref{sec:ignition} yields the bounds on vector and Higgs portal dark matter models, shown in Figure \ref{explmodv} and \ref{explmods} respectively. We fix $\alpha_{X}=0.1$ and $\alpha_\phi = 0.1$ and the dark matter mass, then solve for the value of $\epsilon_{\gamma}$ or $\epsilon_{h}$ corresponding to the per-nucleon cross-section that would cause the 3 Gyr old white dwarf SDSS J160420.40+055542.3 \cite{2017ASPC}, to explode. 

We note that both of the above models imply dark-matter self-interactions, mediated by the Yukawa couplings of the vector and scalar mediators, $i.e.$ the same mediators ($V_{\mu}$ or $\phi$) that couples dark matter to the visible sector. For a scalar $\phi$, this self-interaction is purely attractive. Alternatively, for a vector $\phi$, $XX$ and $\bar{X}\bar{X}$ interactions are repulsive, while $X\bar{X}$ interactions are attractive. A detailed analysis of the self-interaction cross section for this potential, both in the perturbative and non-perturbative regime is given in \cite{Tulin:2013teo}. In Section \ref{sec:ignition} we studied how strongly self-interacting dark matter might affect whether the dark matter sphere at the center of a white dwarf collapses. We found that, even assuming the dark matter is self-thermalized and has the sound speed of an ideal gas (which would correspond to dark matter with substantial self-interactions), a self-gravitating dark matter sphere will satisfy the Jeans instability criterion and collapse at the center of a white dwarf. Therefore, sizable dark matter self-interactions should not substantially alter white dwarf explosion bounds.

\section{Conclusions}
\label{sec:conc}

We have elaborated on a mechanism proposed in \cite{Bramante:2015cua}, whereby non-binary sub-Chandrasekhar white dwarfs can explode as type Ia supernova, through the collapse of heavy asymmetric dark matter in their interiors. A number of new ways the dark matter can ignite white dwarfs by collapsing in their interiors have been investigated, and new bounds were obtained on heavy asymmetric dark matter interactions with nuclei using an old massive white dwarf in the Milky Way. It was found that some very heavy dark matter ($m_X \gtrsim 10^{10}$ GeV) models are unable to trigger a supernova event during collapse, as the required nucleon scattering cross section required is too large given present experimental bounds. However, we have found that such heavy dark matter is still capable of igniting the dwarf, by first collapsing to a black hole, which in turn heats the white dwarf to thermonuclear ignition temperatures through Hawking radiation. In addition, dark matter with a cross section below the minimum for Type Ia ignition via elastic nuclear scattering may collapse forming a black hole. 

Bounds on dark matter-nucleon scattering cross section from asymmetric dark matter igniting white dwarfs can be divided into three dark matter mass ranges. In each of them, a different physical process determines what dark matter nucleon scattering cross section causes white dwarfs to explode. In this work, we also incorporated for the first time a structure factor accounting for the crystalline arrangement of nuclei inside the white dwarf, which hinders dark matter energy loss in dark matter nuclear scattering. This delays substantially the thermalization and collapse of the dark matter sphere, and renders both processes as equal limiting factors for white dwarf ignition at high dark matter masses.

For masses $m_{X}\lesssim{10^{8} \ \gev}$ mass dark matter, the limiting process is the collection of enough dark matter into the white dwarf, so that it forms a sphere which collapses under its own weight at the center of the white dwarf. Here, we have improved on prior work \cite{Bramante:2015cua} in this mass range, by incorporating the effect of multiscatter capture of dark matter on white dwarfs \cite{Bramante:2017xlb}, which we found reduces the cross section required for dark matter to ignite white dwarfs by up to an order of magnitude. 
The white dwarf ignition process during dark matter collapse starts requiring a larger dark matter-nucleon cross section for masses $m_{X}\gtrsim{10^{8} \ \gev}$. This is because for higher dark matter masses, the total mass of the collapsing dark matter sphere is smaller, meaning more interactions between dark matter particles and nuclei are required to prompt white dwarf ignition during collapse. 

For dark matter heavier than $m_{X}\sim 10^{11} \ \gev$, we have determined how the asymmetric dark matter collected inside the white dwarf can spark ignition by first collapsing to a black hole. If the resulting black hole mass is sufficiently small, then radiation dominates over the accretion of stellar matter. The strongly-interacting particles emitted by the black hole have a rather small showering length in the white dwarf material. Upon emission, they undergo rapid inelastic collisions with carbon and oxygen nuclei, generating a shower that travels $\lesssim 10^{-5}$ cm through the rest of the sample. This constitutes an efficient way of transferring the energy lost by the black hole to the white dwarf's core. By the same principles that apply to white dwarf ignition during dark matter collapse, the white dwarf material acquires an homogeneous temperature profile as it is heated.
We find the energy transfer rate to be adequate for ignition by any black hole evaporating with a starting mass $\lesssim 10^{35} \ \gev$. For dark matter to form such black holes in the white dwarf, it must have a large enough dark matter-nucleon cross section for the dark matter sphere to collapse within $\sim{3}$ Gyrs. In addition, we found that dark matter lighter than $m_{X}\lesssim 10^{12} \ \gev$, with a cross section below the necessary for ignition, can form a black hole, which consumes the star without leading to a supernova, so long as the dark matter-nucleon cross section is high enough prevent the black hole from evaporating. In this case, we find both the capture rate and the dark matter thermalization to be the limiting processes when setting bounds on dark matter interactions, based on the existence of an old white dwarf that has not exploded. 

Finally, we have considered here two explicit model for asymmetric, fermionic dark matter with Higgs portal and vector portal couplings to the Standard Model. The vector portal model, in particular, allows for a much faster energy loss rate of dark matter by scattering off electrons in the white dwarf. Bounds on these models were obtained by restricting the nucleon scattering cross section, so that these dark matter models would not have exploded an old white dwarf in the Milky Way. In future work, it will be interesting to consider how the heavy dark matter models presented in this paper and the black hole evaporation ignition mechanism bear on dark matter explanations \cite{Bramante:2015cua} of the Type Ia progenitor problem \cite{Maoz:2013hna}.
\\

{\em Note added:} Shortly after this work appeared on the arxiv, another paper appeared \cite{Janish:2019nkk} with some overlap, which identified a viscous dark matter thermalization regime. This version of our article has been updated to account for viscous thermalization, as discussed in Sec.~\ref{sec:capture}. For dark matter masses in excess of $10^{10}$ GeV, we believe our results differ from \cite{Janish:2019nkk}, because we have accounted for white dwarf structural suppression of dark matter-nuclear scattering at low momentum transfers, as given by Eq.~\eqref{eq:struct} and surrounding text.

\appendix

\section{White Dwarfs}
\label{app:wds}

\begin{figure}[t!]
\includegraphics[scale=0.7]{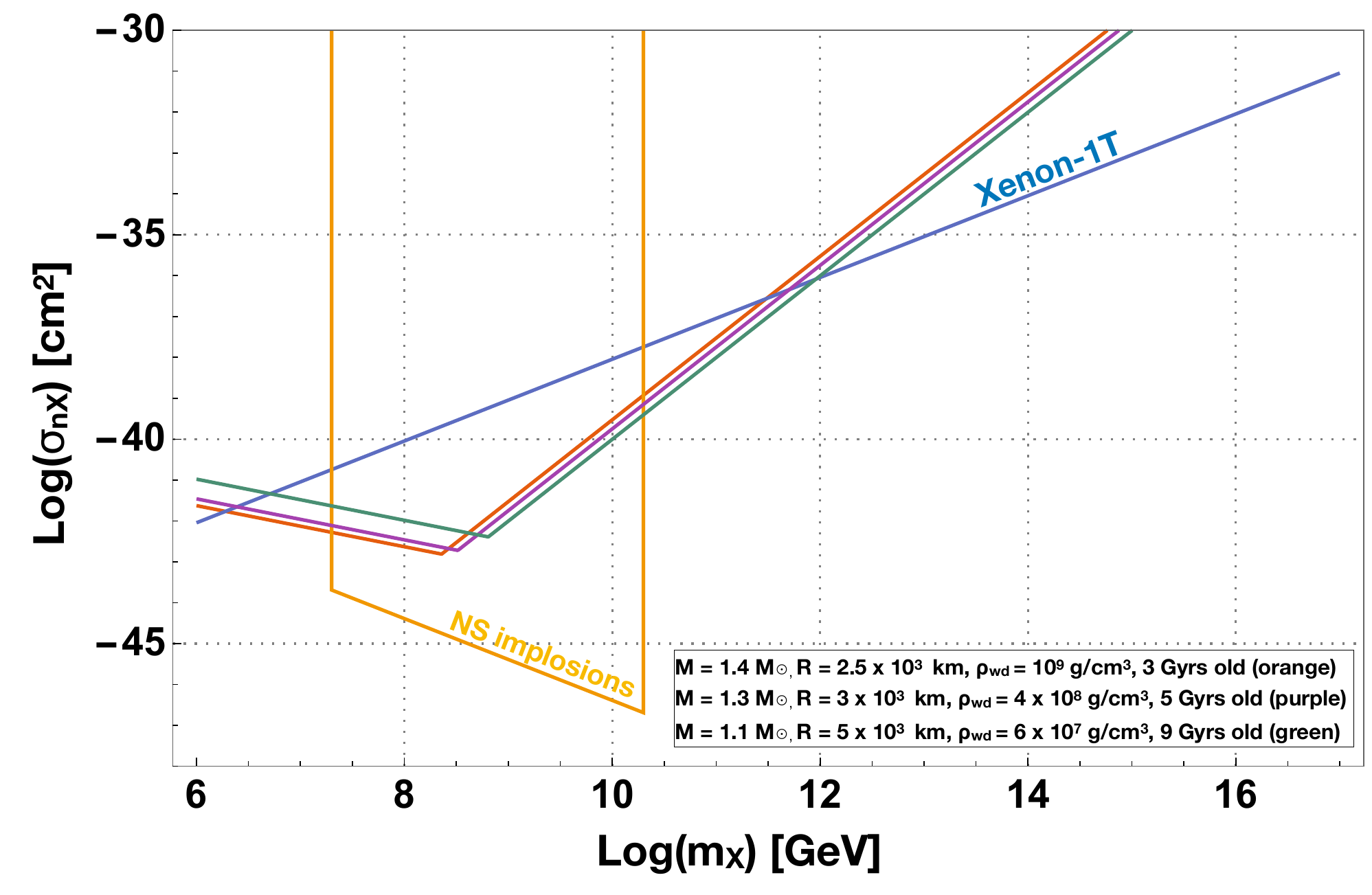} 
\caption{Constraints on the dark matter-nucleon scattering cross-section, as in Figure \ref{fig:bounds2}. Here we show the total parameter space excluded by observation of three different old white dwarfs taken from the Montreal White Dwarf Database \cite{2017ASPC}: SDSS J160420.40+055542.3 has a mass $\sim 1.4M_\odot$ and is $\sim 3 ~{\rm Gyr}$ old, SDSS J073011.24+164800.0 has a mass $\sim 1.3M_\odot$ and is $\sim 5 ~{\rm Gyr}$ old; and SDSS J134113.16+010027.8 has a mass of $\sim 1.1M_\odot$ and is $\sim 9 ~{\rm Gyr}$ old. For dark matter masses up to $m_X \approx 10^8 \ \gev$, the bound on cross-section is determined by the requirement that dark matter collects into a self-gravitating mass at the core of the white dwarf. For larger dark matter masses ($m_X \gtrsim 10^8$ GeV), the bound depends only on the age of the white dwarf, and is relatively insensitive to the white dwarfs composition, see text for details.}
\label{fig:otherwds}
\end{figure}

In this appendix we discuss white dwarf stars that can be used for setting bounds on dark matter interactions with Standard Model particles. In Figure \ref{fig:bounds2}, we presented a bound using a 1.4 $M_\odot$ mass white dwarf that is $\sim 3 ~{\rm Gyr}$ old, found in the Montreal White Dwarf Database \cite{2017ASPC}. This is the oldest $\sim 1.4$ $M_\odot$ mass white dwarf presently observed. Having considered bounds from other old white dwarfs at lower masses, we believe this white dwarf places the most stringent bound available on dark matter-nucleon scattering parameter space, for dark matter masses $\lesssim 10^8$ GeV. The reason for this is as follows: for dark matter masses $m_X \lesssim 10^8$ GeV, the limiting dynamical process by which dark matter ignites white dwarfs, is the collection of a self-gravitating mass of dark matter at the core of the white dwarf over its lifetime. The most massive white dwarfs collect dark matter the fastest, see $e.g.$ \cite{Bramante:2015cua} and Section \ref{sec:capture} of this paper, where it is shown that dark matter capture scales roughly linearly with white dwarf mass, while the required self-gravitating mass decreases with increasing white dwarf mass. The result is that even though the oldest 1.4 $M_\odot$ mass white dwarf is only $\sim 3 ~{\rm Gyr}$ old, compared to older less massive white dwarfs, nevertheless this white dwarf sets the most restrictive bound on the dark matter nucleon cross section, as shown in Figure \ref{fig:otherwds}. 
For dark matter masses greater than $10^8$ GeV, the bound is determined by requiring the dark matter lose energy at the center of the white dwarf rapidly enough that collapse occurs within the lifetime of the white dwarf. It turns out that these collapse dynamics are insensitive to the white dwarf's interior density, since while increasing the density increases the number of scattering targets linearly, on the other hand, in the regime of low momentum transfer (which limits the collapse process, see Section \ref{sec:capture}), the structure factor suppression given in Eq.~\eqref{eq:S''}, scales inversely with white dwarf density. Therefore, for a fixed white dwarf lifetime, the bound on dark matter-nucleon scattering cross-sections is fixed, as shown in Figure \ref{fig:otherwds}.

Next we discuss old white dwarfs in the globular cluster Messier 4 (M4). Pioneering work in \cite{McCullough:2010ai} showed that, assuming that M4 formed inside a dark matter halo which was afterwards stripped during M4's passage through the Milky Way, white dwarfs in near the core of M4 would reside in a $\sim 10^3~{\rm GeV/cm^3}$ background dark matter density. For dark matter masses $\lesssim 10^8$ GeV, this would potentially extend the white dwarf bounds shown in Figure \ref{fig:otherwds} by three orders of magnitude. As detailed in \cite{McCullough:2010ai}, whether or not M4 and other globular clusters formed around a dark matter halo at high redshift is still an active topic of research; see also \cite{Capela:2013yf} for further discussion of this point. Once the provenance of M4 is better established, future work might fruitfully apply the white dwarf ignition dynamics here to the M4 vis-a-vis the methods detailed in Reference \cite{McCullough:2010ai}, and obtain stronger bounds than those presented here. However, for the reasons detailed above, we expect these bounds will mostly apply to dark matter masses $m_X \lesssim 10^8$ GeV.

\section*{Acknowledgments}
We thank J. MacGibbon and J. Niemeyer for useful conversations. Research at Perimeter Institute is supported by the Government of Canada through Industry Canada and by the Province of Ontario through the Ministry of Economic Development \& Innovation. J.~A. and J.~B. acknowledge the support of the Natural Sciences and Engineering Research Council of Canada. J.~B.~thanks the Aspen Center for Physics, which is supported by NSF Grant No.~PHY-1066293.
\bibliographystyle{JHEP}


\bibliography{cdm}

\end{document}